\documentclass[aps,pra,twocolumn,showpacs,superscriptaddress]{revtex4} 
\usepackage{graphicx}
\usepackage{amsmath}
\usepackage{amssymb}
\usepackage{epstopdf}
\usepackage{amsfonts}
\usepackage{dcolumn}
\usepackage{bigints}
\usepackage{xcolor}
\usepackage{hyperref}
\usepackage[normalem]{ulem}
\begin{document}
\title{Two emitters coupled to a bath with Kerr-like non-linearity:
Exponential decay, fractional populations, and Rabi oscillations}
\author{J. Talukdar}
\address{Homer L. Dodge Department of Physics and Astronomy,
  The University of Oklahoma,
  440 W. Brooks Street,
  Norman,
Oklahoma 73019, USA}
\address{Center for Quantum Research and Technology,
  The University of Oklahoma,
  440 W. Brooks Street,
  Norman,
Oklahoma 73019, USA}
\author{D. Blume}
\address{Homer L. Dodge Department of Physics and Astronomy,
  The University of Oklahoma,
  440 W. Brooks Street,
  Norman,
Oklahoma 73019, USA}
\address{Center for Quantum Research and Technology,
  The University of Oklahoma,
  440 W. Brooks Street,
  Norman,
Oklahoma 73019, USA}
\date{\today}

\begin{abstract}
We consider two non-interacting two-level emitters that are coupled weakly to
a one-dimensional non-linear wave guide. 
Due to the Kerr-like non-linearity, the wave guide considered supports---in addition to the scattering 
continuum---a two-body bound state.
As such, the wave guide models a bath with non-trivial mode structure.
Solving the time-dependent  Schr\"odinger equation, the radiation dynamics of the two emitters, 
initially prepared in their excited states, is presented. 
Changing the emitter frequency such that the two-emitter energy is in resonance with one
of the two-body bound states,
radiation dynamics ranging from exponential decay to fractional populations to Rabi oscillations is observed.
Along with the detuning, the dependence on the separation of the two emitters is investigated.
Approximate reduced Hilbert space formulations, which result in effective emitter-separation and momentum
dependent interactions, elucidate the underlying physical mechanisms and provide an avenue to showcase
the features that would be absent if the one-dimensional wave guide did not contain a non-linearity.
Our theoretical findings apply to a number of experimental platforms and the predictions can be tested with state-of-the-art technology. 
In addition, the weak-coupling Schr\"odinger equation based results provide 
critical guidance for the development of master equation approaches.
\end{abstract}
\maketitle

\section{Introduction}
\label{sec_introduction}
Since it is hard to fully isolate quantum systems in realistic experimental settings, 
the quantum mechanical treatment of a system coupled to a bath is important from a practical point of view~\cite{ref_open-review}.
Taking a somewhat philosophical viewpoint, one may furthermore argue that truly isolated quantum systems never exist
since
any quantum system is part of a larger 
universe, i.e., embedded into an environment~\cite{ref_open-quant}. 
In addition, any measurement on the system
involves, according to measurement theory, 
interactions between the system and the environment or bath~\cite{ref_guarnieri,ref_Schlosshauer1,ref_Schlosshauer2}.

The fact that systems are interacting with or can be made to interact with the environment 
that they are embedded into provides a
wealth of opportunities. For example, bath engineering can be used to control the dynamics of
the system, thereby providing an alternative approach to the preparation of pre-specified
target states~\cite{ref_barreiro,ref_diehl,ref_yanay,ref_cian,ref_botzung,ref_zapletal,ref_tomadin,ref_bardyn,ref_lemeshko}. The idea is quite simple. 
As an example, imagine two non-interacting few-level emitters
that are both coupled to a bath. Even though the emitters are not interacting, the 
action of the bath on the emitters can be interpreted as an effective interaction between the emitters.
The effective emitter-emitter interaction can be adjusted, by modifying the mode structure of the bath, 
such that the emitters are driven into a quasi-stationary 
state.

In most cases, the full quantum mechanical treatment of the dynamics of the
entire system, i.e., the system
and the environment, is extremely challenging due to the tremendously large Hilbert
space. 
To make progress, a range of approaches has been pursued~\cite{ref_weimer}. In the weak-coupling limit,
perturbative and master equation approaches have been developed~\cite{ref_rabl_atom-field,ref_pcw}. The strong-coupling
limit can in some cases also be tackled perturbatively~\cite{ref_henriet,ref_wolf,ref_hwang,ref_altintas,ref_mahmoodian,ref_kockum,ref_diaz}.
The present work does not make any of these approximations
and instead analyzes the dynamics of the emitter-waveguide system using the time-dependent
Schr\"odinger equation, working---as, e.g., Refs.~\cite{ref_ripoll-1,ref_ripoll-2,ref_Camacho,ref_Hurst,ref_Piasotski,ref_Dinc,ref_rabl_non-linear}---with the essentially full Hilbert space; the Hilbert space truncation made 
(i.e., dropping of two-photon scattering states) 
leaves the dynamics essentially unchanged for the parameter combinations investigated.
To make the calculations feasible, we restrict ourselves to a one-dimensional
bath with a non-trivial but still relatively simple mode structure. Losses to the ``outside world" 
are neglected entirely, i.e., the emitter-bath 
system is treated as a closed system (the wave guide is assumed to be lossless). 

For concreteness, our work focuses on a photonic lattice with lattice
spacing $a$, nearest neighbor tunneling $J$, and 
on-site interaction $U$~\cite{ref_rabl_non-linear,ref_jugal}. The two two-level emitters are assumed to be located at or
coupled to specific lattice sites
[see Fig.~\ref{fig0}(a)]. 
When both emitters are coupled to the same lattice site, the spacing $x$ vanishes;
when both emitters are coupled to adjacent lattice sites, $x$ is equal to $a$;
and so on. 
This model Hamiltonian was introduced in Ref.~\cite{ref_rabl_non-linear}. While our work builds on the theory framework
introduced in Ref.~\cite{ref_rabl_non-linear},
the emphasis of our study is distinct. Specifically, our work complements Ref.~\cite{ref_rabl_non-linear} in
that we focus on the physics near the bottom or the top of the band as opposed to on the physics in the middle of the band,
i.e., we consider a different range of detuning $\delta/J$, where
$\delta/J$ is defined with respect to the bottom of the two-photon bound state band for negative $U$ and with respect to the top of the two-photon bound state band for positive $U$~\cite{footnote_detuning},
\begin{eqnarray}
\delta =2 \hbar \omega_e - 2 \hbar \omega_c-\text{sign}(U)\sqrt{U^2+16J^2}.
\end{eqnarray}
Here, $\hbar \omega_e$ is the transition energy  of the emitter and $2\hbar \omega_c+\text{sign}(U)\sqrt{U^2+16J^2}$ the bottom of the band for negative $U$ and the top of the band for positive $U$
($\hbar \omega_c$ is the energy  in the middle of the single-photon energy band; see Fig.~\ref{fig1}
for an illustration).
We initialize the emitters in their excited state
$| e\rangle_1 |e\rangle_2$
and the waveguide in the
vacuum state $| \mbox{vac} \rangle$ at time $t=0$. Working in the 
subspace of two excitations, we study the radiation dynamics.
Throughout, we refer to the bath as photon bath. We emphasize, however, that the 
formalism applies also to a phonon bath and baths consisting of other quasi-particles.
The Hamiltonian considered conserves
the total number of excitations 
(see, e.g., Refs.~\cite{ref_ripoll-1,ref_ripoll-2, ref_rabl_non-linear,ref_shi,ref_rabl_atom-field}), 
which is defined as the sum of 
the number of emitter excitations and the number of
photons.
Key objectives of our work are to unravel the dependence of
the
radiation dynamics on the emitter separation $x$ and the detuning $\delta$.

Our main results can be summarized as follows:
(i) The radiation dynamics depends strongly on the emitter separation,
detuning, and strength of the non-linearity.
(ii) Focusing on parameter combinations where the single-photon contributions
can be eliminated adiabatically (this implies moderate $x/a$, not too large $|U|/J$, and detuning $\delta$
such that the system is on resonance with the two-photon band or just slightly off-resonance),
we observe radiation dynamics ranging from exponential decay to fractional population to
Rabi oscillations.
(iii) As discussed in Ref.~\cite{ref_rabl_non-linear}, the Markov approximation provides a faithful description of the exponential decay of the initial state;
our semi-analytic expression for the decay constant is compared with that for a single emitter 
case where the emitter energy is in resonance with the single-photon energy band, excluding the region near the band edge.
(iv) When the onsite interaction is negative and the detuning is chosen such that the energy of the
two emitters is in the band but close to the bottom of the band 
(the actual value of $\delta/J$ depends on the separation $x$ and the coupling strength $g/J$), 
the emitters do not decay to the ground state but instead approach
a steady state that is characterized by
fractional populations. Some of the time-dependent characteristics can be explained
in terms of effective photon-pair--photon-pair interactions.
(v) Detuning extremely close to the bottom of the band 
[as in (iv), the actual value of $\delta/J$ depends on the separation $x$ and the coupling strength $g/J$] 
leads to weakly-damped or essentially undamped Rabi oscillations, which display a notable separation dependence and can be explained in terms of two bound hybridized photonic polaron--excited emitter states.
An analytical two-state model that provides a semi-quantitative description of the Rabi oscillations is derived.  
We note in passing that our results in support of the conclusions summarized under 
(iii) form the basis for developing master equation formulations.

The remainder of this article is organized as follows.
Section~\ref{sec_theory} introduces the model Hamiltonian and 
the
approaches 
used
to solving the time-dependent and time-independent Schr\"odinger equation.
Sections~\ref{sec_results_static} and \ref{sec_results} present our time-independent and time-dependent results.
Last, Sec.~\ref{sec_conclusion} provides a summary and outlook.
Appendix~\ref{appendix_single_emitter} reviews single-emitter results from the literature while
Appendix~\ref{appendixA} contains technical details related to the adiabatic elimination.

\section{System Hamiltonian and theoretical techniques}
\label{sec_theory}

Sections~\ref{sec_hamiltonian} and \ref{sec_ham_bath} 
introduce the full system Hamiltonian and the bath Hamiltonian, respectively.
Our approach for solving the full Schr\"odinger equation is summarized in Sec.~\ref{sec_timedependent}.
The adiabatic elimination of the single-photon states is discussed in Sec.~\ref{sec_adiabatic_elimination}.
Building on the reduced Hilbert space Hamiltonian that results after the adiabatic elimination,
Sec.~\ref{sec_markov_approximation}  discusses the Markov approximation.

\subsection{System Hamiltonian}
\label{sec_hamiltonian}
\begin{figure}[t]
\includegraphics[width=0.49\textwidth]{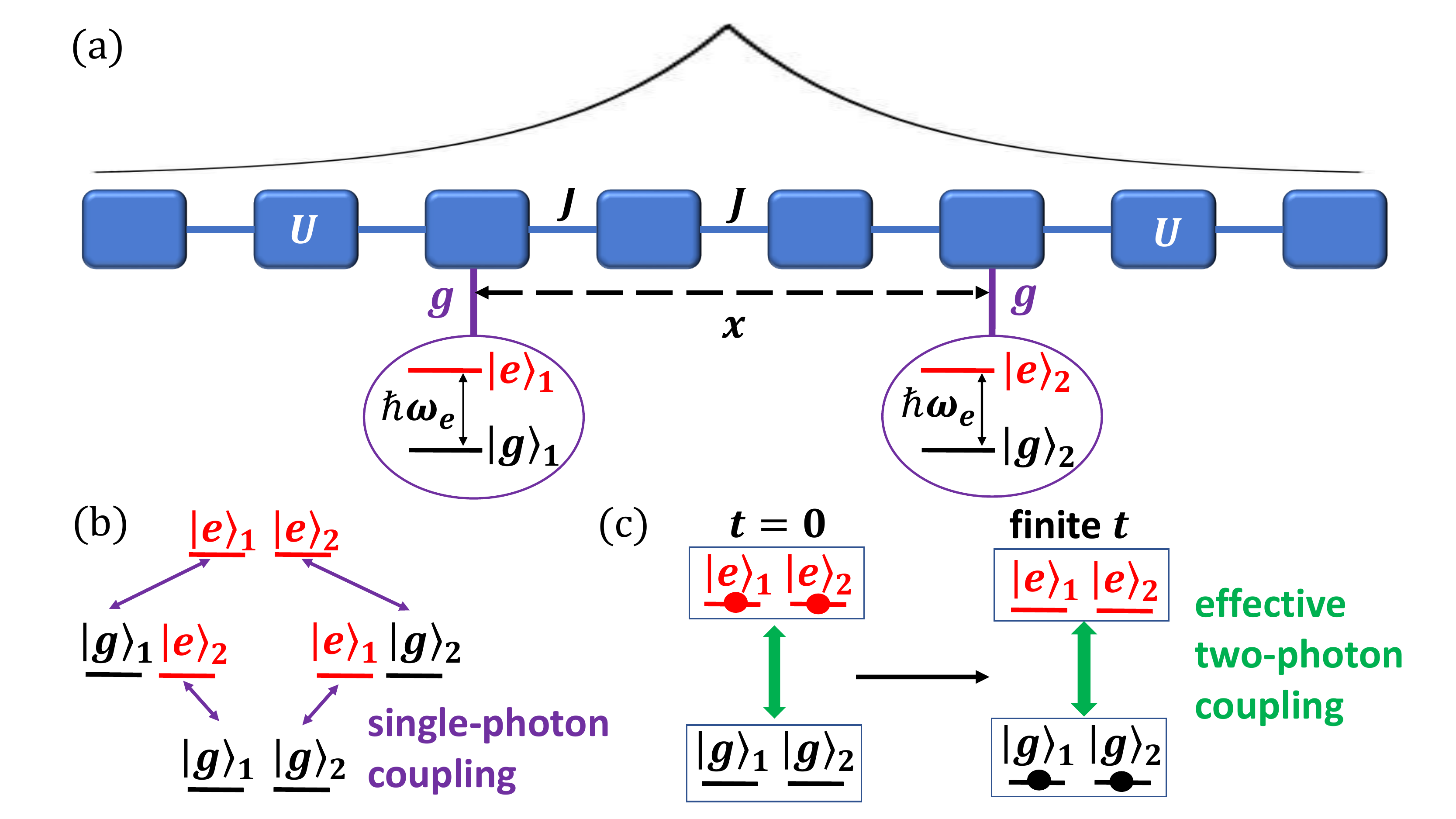}
\caption{Illustration of system under study.
(a) Schematic of system Hamiltonian. Each blue box represents a cavity. The tunnel-coupling between neighboring cavities, which are separated by $a$, is shown by the lines  labeled by $J$. The onsite interaction $U$ characterizes the effective photon-photon interaction; the $U$-term of the Hamiltonian 
$\hat{H}$, Eq.~(\ref{eq_ham_total}), only plays a role for $N_{\text{exc}} \ge 2$.
Two two-level emitters with energy levels $|g\rangle_j$ and $|e\rangle_j$ ($j=1$ or $2$) are coupled to cavities $n_1$ and $n_2$ ($n_1$ and $n_2$ are fixed,
$n_1-n_2=x/a$) with strength $g$. The black line illustrates a two-photon bound state that is supported by the cavity array. The physics explored in this paper occurs in the regime where the size of the two-photon bound state is comparable to the emitter separation $x$.
(b) Illustration of the system-bath Hamiltonian $\hat{H}_{\text{sb}}$, Eq.~(\ref{eq_ham_sb}), in the emitter Hilbert space.
Going from $|e\rangle_1|e\rangle_2|\text{vac}\rangle$ to $|g\rangle_1|g\rangle_2|K\rangle$ requires two
single-photon processes of strength $g$.
(c) Illustration of the effective Hamiltonian $\hat{H}^{\text{adia}}$, Eq.~(\ref{eq_hamiltonian_adia}), in the emitter Hilbert space. The adiabatic elimination introduces an effective two-photon coupling between states $|e\rangle_1|e\rangle_2|\text{vac}\rangle$ and $|g\rangle_1|g\rangle_2|K\rangle$.
 This work monitors the
change of the population of the state $|e\rangle_1|e\rangle_2|\text{vac}\rangle$ with time.}
\label{fig0}
\end{figure}

The total Hamiltonian $\hat{H}$ is given by~\cite{ref_rabl_non-linear,ref_jugal}
\begin{eqnarray}
\label{eq_ham_total}
\hat{H} = \hat{H}_{\text{s}} + \hat{H}_{\text{b}} + \hat{H}_{\text{sb}},
\end{eqnarray}
where $\hat{H}_{\text{s}}$ denotes the system Hamiltonian,
$\hat{H}_{\text{b}}$ the bath Hamiltonian, and
$\hat{H}_{\text{sb}}$ the system-bath coupling
[see Fig.~\ref{fig0}(a) for a schematic].
We consider a system consisting of $N_e$ two-level emitters with
energy separation
$\hbar \omega_e$ between the  ground state $|g\rangle_j$ 
and the excited state $|e\rangle_j$
of the $j$th emitter.
Specifically, $\hat{H}_{\text{s}}$ is given by
\begin{eqnarray}
\label{eq_ham_sys}
\hat{H}_{\text{s}}=
\frac{\hbar \omega_e}{2}
\sum_{j=1}^{N_e} 
(\hat{\sigma}_{j}^z + \hat{I}_j),
\end{eqnarray}
where 
$\hat{\sigma}_{j}^z=|e \rangle _j \langle e| - |g \rangle _j \langle g|$ and $\hat{I}_j=|g \rangle _j \langle g| + |e \rangle _j \langle e|$.
The inclusion of the identity $\sum_{j=1}^{N_e}\hat{I}_j$ in Eq.~(\ref{eq_ham_sys})
introduces an energy shift such that the energy of the 
state with $N_e$ emitters in their excited state and the bath in the vacuum state is equal to $N_e \hbar \omega_e$. The energy shift due to the identity introduces an overall phase in the time-dependent wave packet but does not impact the population dynamics.
The $j$th emitter is coupled to the $n_j$th lattice site of the wave guide, i.e., the emitters
do not move during the dynamics.

Triggered by the system-bath Hamiltonain 
$\hat{H}_{\text{sb}}$ with coupling  strength $g$, the emitters can change their
state from $|e \rangle_j$ to $|g \rangle_j$ 
and from
$|g \rangle_j$ to $|e \rangle_j$,
\begin{eqnarray}\label{eq_ham_sb}
\hat{H}_{\text{sb}}= 
g \sum_{j=1}^{N_e}
\left(
\hat{a}_{n_j} \hat{\sigma}_j^+ + \hat{a}_{n_j}^{\dagger}  \hat{\sigma}_j^-
\right).
\end{eqnarray}
Here, $\hat{\sigma}_{j}^+$ and $\hat{\sigma}_j^{-}$
denote raising and lowering operators of the $j$th emitter,
$\hat{\sigma}_{j}^+ = |e\rangle _j \langle g|$
and
$\hat{\sigma}_{j}^- = |g\rangle _j \langle e|$.
The operators $\hat{a}_{n_j}^{\dagger}$ and $\hat{a}_{n_j}$, respectively, create and destroy
a photon at lattice site $n_j$, where the label $n_j$ takes values from $1$
to $N$ with $N$ denoting the number of lattice
sites or cavities of the wave guide. We are interested in the regime where the dynamics is independent of $N$
(large $N$ limit). 
Since the system-bath Hamiltonian does not include any counterrotating terms, the treatment is restricted to the weak-coupling regime where 
$|g|$ is small compared to the other energy scales of the system~\cite{ref_petro-book}.

The Hamiltonian $\hat{H}_{\text{b}}$ 
is taken to be a one-dimensional array of tunnel coupled cavities
in the tight-binding limit.
It
is characterized by the ``photon energy" $\hbar \omega_c$ (the middle of the single-photon
energy band has energy $\hbar \omega_c$),
the tunneling energy $J$, and the onsite interaction energy $U$:
\begin{eqnarray}
\label{eq_bath_ham}
\hat{H}_{\text{b}} =&&
\hbar \omega_c  \sum_{n=1}^{N} 
 \hat{a}_n^{\dagger} \hat{a}_n 
-J  \sum_{n=1}^{N} \left(
\hat{a}_n^{\dagger} \hat{a}_{n+1} + \hat{a}_{n+1}^{\dagger} \hat{a}_{n} 
\right) \nonumber \\
&&+\frac{U}{2} 
\sum_{n=1}^{N}
 \hat{a}_n^{\dagger}\hat{a}_n^{\dagger}\hat{a}_n\hat{a}_n .
\end{eqnarray}
 In Eq.~(\ref{eq_bath_ham}), the 
photons are assumed to interact, due to the presence of a
Kerr-like medium, either effectively repulsively ($U>0$) or effectively attractively  
($U<0$). A positive $U$ gives rise to a two-photon bound state 
with center-of-mass wave vector $K$
that lies above the two-photon
continuum while a negative $U$ gives rise to a two-photon bound state with center-of-mass wave vector $K$
that lies below the
two-photon scattering continuum~\cite{ref_molmer1, ref_molmer2,ref_valiente, ref_petrosyan}. 

The bath Hamiltonian considered here 
has been chosen for several reasons:
(i) The eigen energies and eigen states of $\hat{H}_{\text{b}}$ are known analytically (see below)~\cite{ref_valiente}.
(ii) Despite its simplicity, the Hamiltonian $\hat{H}_{\text{b}}$ supports a non-trivial mode structure, namely
the above-mentioned two-photon bound state~\cite{ref_molmer1, ref_molmer2,ref_valiente, ref_petrosyan}.
(iii)  It was predicted in Ref.~\cite{ref_rabl_non-linear}
that the emission dynamics of $\hat{H}$
displays, for certain parameter combinations, sub-radiance and super-correlations.
These intriguing 
findings motivate our quest to map out 
constructive and destructive interferences,  with the goal of identifying the dominant 
emission pathways. 
Throughout, we are interested in situations where the initial $t=0$ state contains two 
excitations in the emitter Hilbert space (i.e., $| \Psi(0) \rangle = | e \rangle_1 | e\rangle_2 |\text{vac} \rangle$)
and
where the dynamics is driven, at least in part, by the 
non-trivial mode structure of the bath, i.e., by the existence of the two-photon bound states supported by $\hat{H}_{\text{b}}$. 
For brevity,
we adopt the notation
$| e \rangle_1 | e\rangle_2 |\text{vac} \rangle=| e, e ,\text{vac} \rangle$, etc.
The next section discusses selected properties of $\hat{H}_{\text{b}}$. 

\begin{figure}[t]
\includegraphics[width=0.33\textwidth]{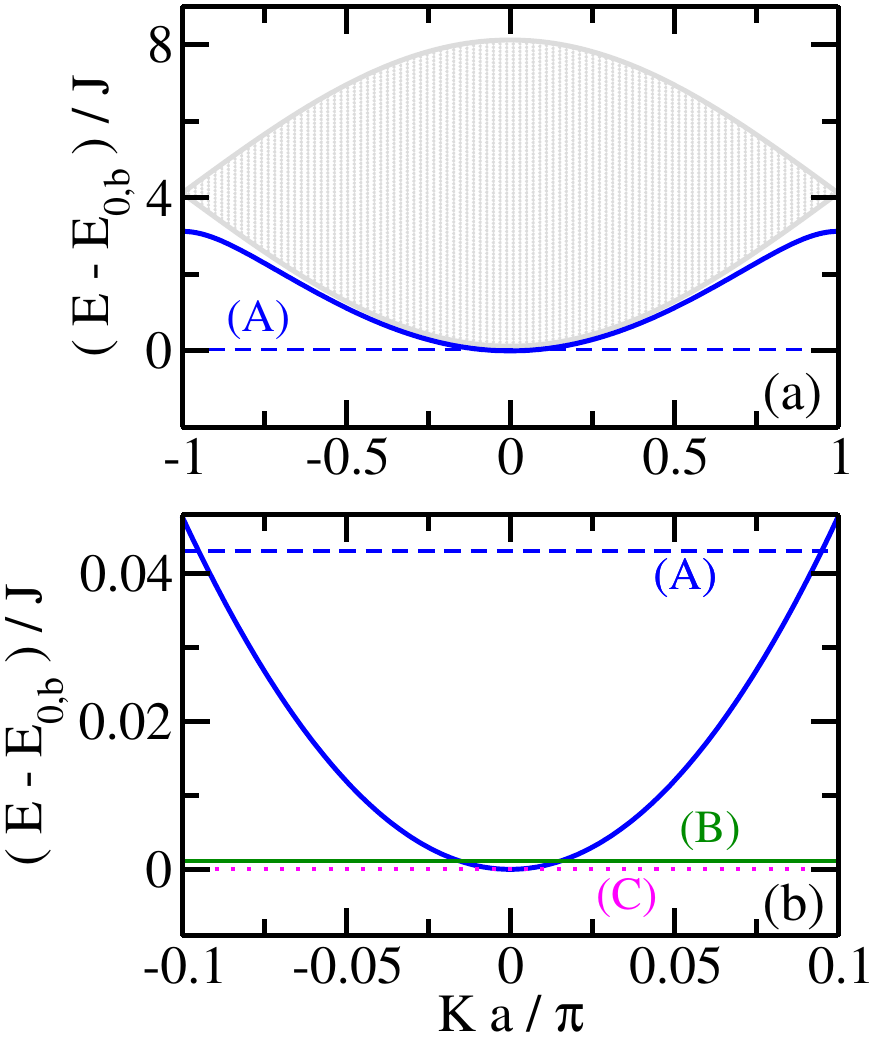}
\caption{Two-photon eigen spectrum as a function of the scaled center-of-mass wave number 
$Ka  / \pi$. 
Note that the energy is shifted such that the bottom of the two-photon bound state band sits at zero.
(a) The gray-shaded energy band corresponds to the two-photon
scattering continuum, 
Eq.~(\ref{eq_twophoton_scatteringenergy}). The thick blue solid line shows the energy $E_{K,b}$ of the two-photon bound state for $U/J=-1$. 
While the gray band and thick blue solid line appear to coincide for $K=0$ on the scale shown, we note that the bottom of the 
two-photon scattering continuum at $K=0$ lies  $(-4+\sqrt{17})J\approx 0.123 J$ above the $K=0$ two-photon bound state energy. This separation is sufficiently large for the two-photon scattering continuum to play a negligible role in the system dynamics considered in this paper.
The thin dashed line 
shows the energy of the state $|e,e , \mbox{vac} \rangle$ for
$\delta/J=0.0431$.
(b) Blow-up of (a), focusing on the region around the bottom of the band.
The blue solid and blue dashed lines are the same as in (a).
The horizontal green solid and magenta dotted lines 
show the energy of the state $|e,e , \mbox{vac} \rangle$ for $\delta/J=0.0011$ and $0.0001$, respectively.
The labels ``(A)", ``(B)", and ``(C)"
refer to the three scenarios
introduced in the second to last paragraph of Sec.~\ref{sec_hamiltonian}.
}
\label{fig1}
\end{figure}

The Hamiltonian $\hat{H}$ has four independent energy scales:
$\delta$, $g$, $J$, and $U$.
Throughout, $J$, $\hbar/J$, and $a$ are used as energy unit, time unit, and
length unit, respectively.
To reduce the parameter space, we 
analyze the system properties for fixed $g/J$ as functions of 
 $\delta/J$ and $x/a$. The dependence on $U/J$ is explored a bit; most calculations
 presented,
 however, are for $U/J=-1$.
 Throughout, we work in the weak coupling regime, i.e., we use $g/J=1/50$.
 Section~\ref{sec_conclusion} comments briefly on the dependence of the system properties on $g/J$.
As illustrated in Fig.~\ref{fig1}, the detuning $\delta$ is set such that 
the energy $2 \hbar \omega_e$ of the initial state is, for $g=0$, 
(A) in resonance with a two-photon bound state 
with $K a/ \pi$ not too close to $0$ and not too close to $\pm 1$
[see the horizontal dashed line
in Fig.~\ref{fig1}(b) as an example];
(B) in resonance with a two-photon bound state 
with $K a/\pi$ 
a bit larger than $0$
[see the horizontal solid line in Fig.~\ref{fig1}(b) as an example];
and
(C) in resonance with the two-photon bound state extremely close to the bottom of the band 
[$K^{(0)}a/ \pi\approx 4.5 \times 10^{-3}$; see the horizontal dotted line
in Fig.~\ref{fig1}(b) as an example].

We note that single-photon losses are not included in our treatment. 
This is justified if the dynamics governed by  $\hat{H}$ 
is notably faster than the dynamics associated with
the single-photon losses.
Using the parameters of Fig.~\ref{fig_radiation_dynamics} as an example, this 
implies that the single-photon loss rate is assumed to be smaller than
$\approx 10^{-4}J/\hbar$.

\subsection{Mode structure of the bath Hamiltonian}
\label{sec_ham_bath}

Since the Hamiltonian $\hat{H}_{\text{b}}$ commutes with the photon number operator
$\hat{N}$~\cite{ref_molmer1,ref_molmer2,ref_valiente},
\begin{eqnarray}
\hat{N}=\sum_{n=1}^{N} \hat{a}_n^{\dagger} \hat{a}_n, 
\end{eqnarray}
$\hat{H}_{\text{b}}$ is block-diagonal in the number of
photons.
In what follows, 
we discuss the eigen spectrum of $\hat{H}_{\text{b}}$ in the one- and two-photon subspaces.

We start with the single-photon subspace.
The single-photon
energy $E_k$ reads~\cite{ref_sakurai}
\begin{eqnarray}
\label{eq_energy_onephoton}
E_k = \hbar \omega_c - 2 J \cos \left( ka \right),
\end{eqnarray}
where the single-photon wave number $k$ ($k a/ \pi \in [-1,1 ]$) is a good quantum
number.
The single-photon eigen states 
with energy $E_k$ are denoted by $|\psi_k \rangle$.
Equation~(\ref{eq_energy_onephoton}) shows that $E_k$
is equal to $\hbar \omega_c$ for $k a/ \pi = \pm 1/2$ (this is the middle of the band),
equal to $\hbar \omega_c - 2J $ for $k a/ \pi = 0$ (this is the bottom of the band), and equal to $\hbar \omega_c + 2J $ for $ka/ \pi = \pm 1$.
For later reference, we note that the single-photon group velocity $v_k$ is given by
\begin{eqnarray}
v_k = \frac{2 J a}{\hbar} \sin(ka).
\end{eqnarray} 
This shows that a single photon travels, ``on average," two lattice sites
per characteristic time $\hbar/J$ for $k a / \pi=\pm 1/2$ and not at all for $k a=0$ and $k a / \pi = \pm 1$.
According to this classical average-speed-picture, two individually launched photons may not interfere with each other 
if the photon's wave number is close to zero or $\pm \pi/a$,
or if the emitters are separated by many lattice sites.

We now turn to the
two-photon subspace, which is spanned by scattering 
states $|\psi_{K,q} \rangle$ with energy $E_{K,q}$ and bound states $|\psi_{K,b} \rangle$ with energy $E_{K,b}$~\cite{ref_molmer1, ref_molmer2,ref_valiente, ref_petrosyan}.
The center-of-mass wave number $K$ 
is a good quantum number.
The gray band in Fig.~\ref{fig1}(a) shows
the two-photon scattering energy $E_{K,q}$~\cite{ref_molmer1,ref_molmer2,ref_valiente},
\begin{eqnarray}
\label{eq_twophoton_scatteringenergy}
E_{K,q} = 2 \hbar \omega_c - 4 J \cos \left( \frac{Ka}{2} \right)
\cos ( qa),
\end{eqnarray}
as a function of  $K$ ($Ka / \pi \in [-1,1]$). 
The energy continuum arises from the fact
that the relative wave number $q$ can take a range of values that depends on $K$
(e.g., $q a / \pi \in [-1,1]$ for $K=0$ and $qa / \pi=0$ for $K a / \pi = \pm 1$).
The middle of the scattering continuum lies at $2 \hbar \omega_c$, and 
the scattering continuum has a width of $8J$ for $Ka / \pi=0$ and a width
of $0$ for $Ka / \pi= \pm 1$.
While the two-photon scattering energies $E_{K,q}$ are independent of $U$, the associated
scattering states depend on $U$.

In addition, the Hamiltonian $\hat{H}_{\text{b}}$ supports
one two-photon bound state with energy $E_{K,b}$ for each $K$~\cite{ref_molmer1, ref_molmer2,ref_valiente, ref_petrosyan},
\begin{eqnarray}
E_{K,b} = 2 \hbar \omega_c + \mbox{sign}(U) \left[ U^2 + 16 J^2 \cos^2 \left( \frac{Ka}{2} \right) \right]^{1/2}.
\end{eqnarray} 
For negative $U/J$, the bound state lies below
the scattering continuum (see the thick blue solid line in Fig.~\ref{fig1} for $U/J=-1$).
In this case,
the binding energy for a given $K$ is defined as the energy difference between the
lower edge of the scattering continuum ($E_{K,q}$ with $q=0$) and the bound state energy $E_{K,b}$.
The situation for positive $U/J$ is similar, except that the bound state lies above
the scattering continuum.
The binding energy increases with increasing $|U|/ J$; correspondingly,
the two-photon bound state wave function becomes more localized.
We note that two-photon bound states~\cite{ref_firstenberg} and 
repulsively bound atom pairs in optical lattices~\cite{ref_rpl-bound}
have been observed experimentally.

The horizontal lines in Fig.~\ref{fig1}(b)
show the energy of the 
state $|e \rangle_1 |e \rangle_2 | \mbox{vac} \rangle$ for three different values of $\delta/J$:
$\delta/J=0.0431$ (dashed line) corresponds to scenario (A),
$\delta/J=0.0011$ (solid line) corresponds to scenario (B), and
$\delta/J=0.0001$ (dotted line) corresponds to scenario (C).
The crossings between the energy of the initial state and the energy $E_{K,b}$ of the two-photon bound state 
define the uncoupled (i.e., $g=0$) resonance wave numbers 
$\pm K^{(0)}$~\cite{ref_rabl_non-linear}, where
$K^{(0)}$ is defined to be positive. 
For finite coupling strength $g$, the value of the resonance wave vector shifts from $K^{(0)}$ to $K^{(*)}$
(see Appendix~\ref{appendixA} for details).

In scenario (A), radiation is emitted predominantly, via intermediate single-photon states, 
into two-photon bound states
with wave numbers $\approx \pm K^{(*)}$, leading to exponential decay~\cite{ref_rabl_non-linear}.
Since the group velocity $v_{K,b}$~\cite{ref_valiente},
\begin{eqnarray}
v_{K,b} = \frac{1}{\hbar} \frac{\partial E_{K,b}}{\partial K},
\end{eqnarray}
of the two-photon bound state
depends on $K$ ($v_{K,b}$ is zero for $Ka / \pi=0$ and $\pm 1$ and finite for 
all other $Ka$), 
the decay constant shows a distinct dependence on the resonance wave number
or, equivalently, on the detuning $\delta/J$~\cite{ref_rabl_non-linear}.
In scenario (B), the  near-flatness of the band implies that the initial energy
is nearly equal to that of several two-photon bound states
with $|Ka / \pi| \ll 1$. This leads, as shown in Sec.~\ref{sec_results},
to fractional populations.
In scenario (C), the two-photon bound state
band splits into a band and an emitter-photon coupling induced polaron-like bound state that
hybridizes with the state $|e,e,\text{vac} \rangle$ upon inclusion of the coupling between
the polaron-like bound state and the state $|e,e,\text{vac} \rangle $, leading to essentially
undamped Rabi oscillations that are reproduced very well by a two-state model.
Selected results for scenarios (B) and (C) are discussed in Ref.~\cite{ref_jugal}.

\subsection{Solving the Schr\"odinger equation}
\label{sec_timedependent}
Since $\hat{H}$ commutes with the 
excitation
operator $\hat{N}_{\text{exc}}$~\cite{ref_ripoll-1,ref_ripoll-2, ref_rabl_non-linear,ref_shi,ref_rabl_atom-field},
\begin{eqnarray}
\hat{N}_{\text{exc}}=
\hat{N}
+ 
\sum_{j=1}^{N_e} \hat{\sigma}^+_j \hat{\sigma}^-_j,
\end{eqnarray}
the number  of excitations $N_{\text{exc}}$ (eigenvalue of $\hat{N}_{\text{exc}}$) is conserved.
Correspondingly, the time evolution 
of an initial state with $N_{\text{exc}}=2$
under the Hamiltonian $\hat{H}$, Eq.~(\ref{eq_ham_total}), can be expanded in terms of the 
states $|e,e,\mbox{vac} \rangle$,
$\hat{a}_n^{\dagger} | e,g,\mbox{vac} \rangle $,
$\hat{a}_n^{\dagger} | g,e,\mbox{vac} \rangle $,
and
$\hat{a}_n^{\dagger} \hat{a}_{n'}^{\dagger}| g,g,\mbox{vac} \rangle $,
where $n$ and $n'$  take the values $1,\cdots,N$.
Alternatively,
the time-dependent state $|\Psi(t) \rangle$ 
can be expanded
using the zero-, one-, and two-photon  eigen states of $\hat{H}_{\text{b}}$~\cite{ref_rabl_non-linear},
\begin{eqnarray}
\label{eq_wavepacket_ansatz}
|\Psi(t) \rangle  = &&
\exp(- 2\imath \omega_e t)
[
c_{ee}(t) |e,e,\mbox{vac} \rangle + \nonumber \\
&&\sum_k  c_{1k}(t) \hat{a}_k^{\dagger} |e,g,\mbox{vac} \rangle + \nonumber \\
&& \sum_k c_{2k}(t) \hat{a}_k^{\dagger} |g,e,\mbox{vac} \rangle + \nonumber \\
&&\sum_K c_{K,b}(t) \hat{P}_{K,b}^{\dagger} |g,g,\mbox{vac}\rangle + \nonumber \\
&&\sum_{K,q} c_{K,q}(t) \hat{P}_{K,q}^{\dagger} |g,g,\mbox{vac}\rangle ], 
\end{eqnarray}
where 
$|\psi_k\rangle=\hat{a}_k^{\dagger}|\mbox{vac} \rangle$,
$|\psi_{K,b} \rangle = \hat{P}_{K,b}^{\dagger}|\mbox{vac} \rangle$, and
$|\psi_{K,q} \rangle = \hat{P}_{K,q}^{\dagger}|\mbox{vac} \rangle$.
The operators $\hat{a}_n^{\dagger}$ and $\hat{a}_k^{\dagger}$ are related
via a Fourier transform in the standard way,
\begin{eqnarray}
\hat{a}_k^{\dagger} = \frac{1}{\sqrt{N}} \sum_{n=1}^{N}\exp(\imath k a n) \hat{a}_n^{\dagger}.
\end{eqnarray}

Inserting Eq.~(\ref{eq_wavepacket_ansatz}) into
the time-dependent Schr\"odinger equation
\begin{eqnarray}
\imath \hbar \frac{\partial \Psi(t)}{\partial t} = \hat{H} \Psi(t)
\end{eqnarray}
and projecting onto the basis states,
we obtain a set of first-order differential equations for the
time-dependent expansion coefficients~\cite{ref_rabl_non-linear},
\begin{eqnarray}
\label{eq_coeffe}
\imath \hbar \dot{c}_{ee}(t) = \frac{g}{\sqrt{N}}\sum_{\alpha=1,2}\sum_k  \exp(\imath k a n_{\beta}) c_{\alpha k}(t) ,
\end{eqnarray}
\begin{widetext}
\begin{eqnarray}
\label{eq_coeffk}
\imath \hbar \dot{c}_{\alpha k}(t) =
\Delta_k
c_{\alpha k}(t)  
+\frac{g}{\sqrt{N}} \exp( -\imath k a n_{\beta}) c_{ee}(t) +
\frac{g}{N} \sum_K M_b(k, n_{\alpha},K) c_{K,b}(t) +
\frac{g}{N} \sum_{K,q} M_q(k, n_{\alpha},K) c_{K,q}(t),
\end{eqnarray}
\begin{eqnarray}
\label{eq_coeffbound}
\imath \hbar \dot{c}_{K,b}(t) = 
\Delta_{K,b}
c_{K,b}(t) + \frac{g}{N} \sum_{\alpha=1,2} \sum_k [M_b(k ,n_{\alpha},K)]^* c_{\alpha k}(t),
\end{eqnarray}
and
\begin{eqnarray}
\label{eq_coeffscatt}
\imath \hbar \dot{c}_{K,q}(t) = 
\Delta_{K,q}
c_{K,q}(t)
+ \frac{g}{N} \sum_{\alpha=1,2} \sum_k  [M_q(k, n_{\alpha},K)]^* c_{\alpha k}(t).
\end{eqnarray}
\end{widetext}
For $N_{\text{exc}}=2$ (recall, this is the focus of our work),
Eqs.~(\ref{eq_coeffe})-(\ref{eq_coeffscatt}) are equivalent to the time-dependent
Schr\"odinger equation.
The quantities $\Delta_k$, $\Delta_{K,b}$, and $\Delta_{K,q}$ denote energy detunings:
\begin{eqnarray}
\Delta_k=E_k-\hbar \omega_e,
\end{eqnarray}
\begin{eqnarray}
\Delta_{K,b}=E_{K,b}-2 \hbar \omega_e,
\end{eqnarray}
and
\begin{eqnarray}
\Delta_{K,q}=E_{K,q}- 2 \hbar \omega_e.
\end{eqnarray}
In Eq.~(\ref{eq_coeffk}), $\alpha$ takes the values $1$ or $2$. The value of $\beta$ depends on $\alpha$:
$\beta=2$ for $\alpha=1$
and 
$\beta=1$ for $\alpha=2$  in Eqs.~(\ref{eq_coeffe})-(\ref{eq_coeffk}).
The matrix elements $M_b(k,n,K)$  and $M_q(k,n,K)$
measure the contribution of a photon
with wave number $k$ to the two-photon bound state and to the two-photon scattering state, respectively,
after acting with $\hat{a}_{n}$ on the two-photon state,
\begin{eqnarray}
M_b(k, n,K)=
N\langle \psi_k|\hat{a}_{n} | \psi_{K,b} \rangle
\end{eqnarray}
and
\begin{eqnarray}
M_q(k, n,K) = N\langle \psi_k | \hat{a}_{n} | \psi_{K,q} \rangle.
\end{eqnarray}
The matrix element   
$M_b(k, n,K)$ reads~\cite{ref_rabl_non-linear}
\begin{eqnarray}
\label{eq_mb_element}
\nonumber M_b(k, n,K)=\sqrt{2}\times \nonumber \\
\sum_m \exp \left[ \imath m \left(k-\frac{K}{2} \right)a 
+ \imath n (K-k)a \right] \psi_{K,b}(m)],
\end{eqnarray}
where $\psi_{K,b}(m)=\langle r=ma|\psi_{K,b} \rangle$ with $|r\rangle$ denoting the
relative distance between the two photons;
$M_q(k, n,K)$ is obtained by replacing the subscript $b$ in Eq.~(\ref{eq_mb_element})
by $q$. 
The matrix elements
$M_b(k, n,K)$ and $M_q(k, n,K)$ are defined such that their values for a given $k$, $n$, and $K$ [and $q$ for $M_q(k, n,K)$] are independent of  $N$; they differ by a factor $N$ from those defined in Ref.~\cite{ref_rabl_non-linear}.

We solve the coupled differential equations by discretizing the wave numbers $k$, $K$, and $q$. 
For $N$ lattice sites and $N_{\text{exc}}=2$ excitations, we have 
$N^2 + (N_e+1)N+1$
expansion coefficients. 
If the scattering continuum can be neglected, the computational complexity reduces dramatically since
the number of coupled equations reduces from order $N^2$ to order $N$.
For the parameters considered in this work, we found---by performing
calculations for $N \le 300$---that the scattering continuum plays a negligible role.
This is consistent with the findings of Ref.~\cite{ref_rabl_non-linear}.
Thus, the results presented are calculated using $N$ up to $9001$, excluding the scattering continuum from the Hilbert space.

Two numerical approaches are used.
First, we use the Runge-Kutta algorithm~\cite{ref_numerical-recipe} with adjustable time step to propagate the coefficients for a given initial 
 state at $t=0$
to time $t$.
Second,  we 
express the Hamiltonian $\hat{H}$ in terms of the uncoupled $g=0$ basis states
using the matrix elements defined above. Determining the finite $g$ eigen states through diagonalization,
we project the initial state onto the eigen states of $\hat{H}$.
Since the exact diagonalization approach is numerically more stable, the results presented in this paper are obtained using that approach.

The eigen spectrum of $\hat{H}$ provides complementary clues for understanding the 
emitter dynamics. For finite $g$, eigen states with hybridized character that contain photon and emitter contributions can exist~\cite{ref_jugal}; 
for certain parameter combinations, these strongly-mixed states have energies that lie ``outside" the two-photon
bound state band.
These states are discussed in more detail in Sec.~\ref{sec_results_static}.
Hybridized light-matter states play a critical role in many other related contexts~\cite{ref_bykov,ref_kofman, ref_john-prl, ref_kimble-1}.

\subsection{Adiabatic elimination of single-photon states}
\label{sec_adiabatic_elimination}
This section discusses the construction of a reduced dimensionality 
Hamiltonian  that ``lives" in the Hilbert space spanned by the states
$|g,g,\mbox{vac} \rangle$, 
$ \hat{P}_{K,b}^{\dagger} | g,g,\mbox{vac} \rangle$, and
$ \hat{P}_{K,q}^{\dagger} | g,g,\mbox{vac} \rangle$.
The basis states $\hat{a}_k^{\dagger} | e,g,\mbox{vac} \rangle$
and $\hat{a}_k^{\dagger} | g,e,\mbox{vac} \rangle$ are removed
and accounted for approximately through effective interactions in the reduced dimensionality
Hilbert space
[see Figs.~\ref{fig0}(b) and \ref{fig0}(c)].
The construction of the reduced dimensionality Hamiltonian is based on the
adiabatic elimination of $c_{1k}(t)$ and $c_{2k}(t)$ from the 
coupled 
equations~\cite{ref_rabl_non-linear,ref_carmichael}.
The approximation requires that
the change of $c_{1k}(t)$ and $c_{2k}(t)$ 
with time in Eqs.~(\ref{eq_coeffe})-(\ref{eq_coeffscatt}) can be neglected.

Even though the adiabatic elimination approach removes $c_{1k}(t)$ and $c_{2k}(t)$
from the coupled equations, 
we emphasize that the single-photon states play an important role even in the regime where the differential equations
after adiabatic elimination provide a faithful description of the dynamics. This can be seen by 
inspecting Eqs.~(\ref{eq_coeffe})-(\ref{eq_coeffscatt}).
If we start, e.g., with $c_{ee}(0)=1$, then the evolution of the initial state for $t=0^+$ is driven by the change of 
$\dot{c}_{\alpha k}(t)$;
this follows since the $c_{ee}(t)$-coefficient appears on the right hand side of Eq.~(\ref{eq_coeffk})
but not on the right hand side of Eqs.~(\ref{eq_coeffbound}) and (\ref{eq_coeffscatt}).
The key point of the adiabatic elimination is that the single-photon states serve
as intermediate states---population goes into and out of these states at roughly equal rates such that
the majority of the population is in the basis states with two emitter excitations and
zero emitter excitations. 
The adiabatic elimination breaks down when $g$ becomes too large
(the actual value of $g/J$ depends on 
the values of $\delta/J$, $U/J$, and $x/a$).

Carrying out the adiabatic elimination 
and neglecting the effective 
coupling matrix elements
$H_{K,K',q}(n_1,n_2)$ and $J_{K,K',q,q'}(n_1,n_2)$ 
that involve the two-photon scattering continuum
(see Appendix~\ref{appendixA}), 
Eqs.~(\ref{eq_coeffe})-(\ref{eq_coeffscatt})
reduce to a set of differential equations
that can be written in terms of the effective adiabatic Hamiltonian $\hat{H}^{\text{adia}}$.
In matrix form, we find~\cite{ref_rabl_non-linear}
\begin{eqnarray}
\label{eq_se_adia1}
\imath \hbar \frac{\partial}{\partial t}
\left(
\begin{array}{c}
c_{ee}(t) \\
\vec{c}_{K,b}(t)
\end{array}
\right)
=
\underline{H}^{\text{adia}} 
\left(
\begin{array}{c}
c_{ee}(t) \\
\vec{c}_{K,b}(t)
\end{array}
\right),
\end{eqnarray}
where
\begin{eqnarray}
\vec{c}_{K,b}(t) = 
(c_{K_1,b}(t),\cdots,c_{K_N,b}(t))^T,
\end{eqnarray}
\begin{widetext}
\begin{eqnarray}
\label{eq_hamiltonian_adia}
\underline{H}^{\text{adia}} 
&=&
\left(
\begin{array}{cc}
2 \Delta_e &0 \\
0 & \underline{\Delta}_{K,b}
\end{array}
\right) +
\frac{g^2}{J} \left(
\begin{array}{cc}
0 & N^{-1/2} (\vec{F}_{K,b}(n_1,n_2)) ^T \\
N^{-1/2} [\vec{F}_{K,b}(n_1,n_2)]^* & N^{-1}\underline{G}_{K,K'}(n_1,n_2)
\end{array}
\right),
\end{eqnarray}
\end{widetext}
and
\begin{eqnarray}
\label{eq_se_adia4}
\vec{F}_{K,b}(n_1,n_2) = 
(F_{K_1,b}(n_1,n_2),\cdots,F_{K_N,b}(n_1,n_2))^T.
\end{eqnarray}
The definition of the vector
$\vec{F}_{K,b}(n_1,n_2)$ is given in Eq.~(\ref{eq_fsubcapk}).
The matrices $\underline{\Delta}_{K,b}$ and 
$\underline{G}_{K,K'}(n_1,n_2)$ have dimension
$N \times N$: $\underline{\Delta}_{K,b}$ is diagonal with
$\Delta_{K_1,b},\cdots,\Delta_{K_N,b}$ on the diagonal
and the elements of $\underline{G}_{K,K'}(n_1,n_2)$ 
are given by $G_{K_l,K_{l'}}(n_1,n_2)$ [Eq.~(\ref{eq_capg})],
with $l$ and $l'$ taking the values $1,\cdots,N$. 
The second term on the right hand side of
Eq.~(\ref{eq_hamiltonian_adia}) represents effective interactions 
that arise due to the elimination of the single-photon states.
The element $F_{K_l,b}(n_1,n_2)$ 
represents an effective interaction between
the state $|e,e,\mbox{vac} \rangle$ and the two-photon bound state with center-of-mass wave number 
$K_l$ while
the element $G_{K_l,K_{l'}}(n_1,n_2)$ represents
an effective interaction between 
the two-photon bound state with wave number $K_l$ and the two-photon bound state with wave number $K_{l'}$.
We note that $F_{K_l}(n_1,n_2)$ and $G_{K_l,K'_{l'}}(n_1,n_2)$ 
are independent of $g$ and,
in general, complex.

\begin{figure}
\includegraphics[width=0.48\textwidth]{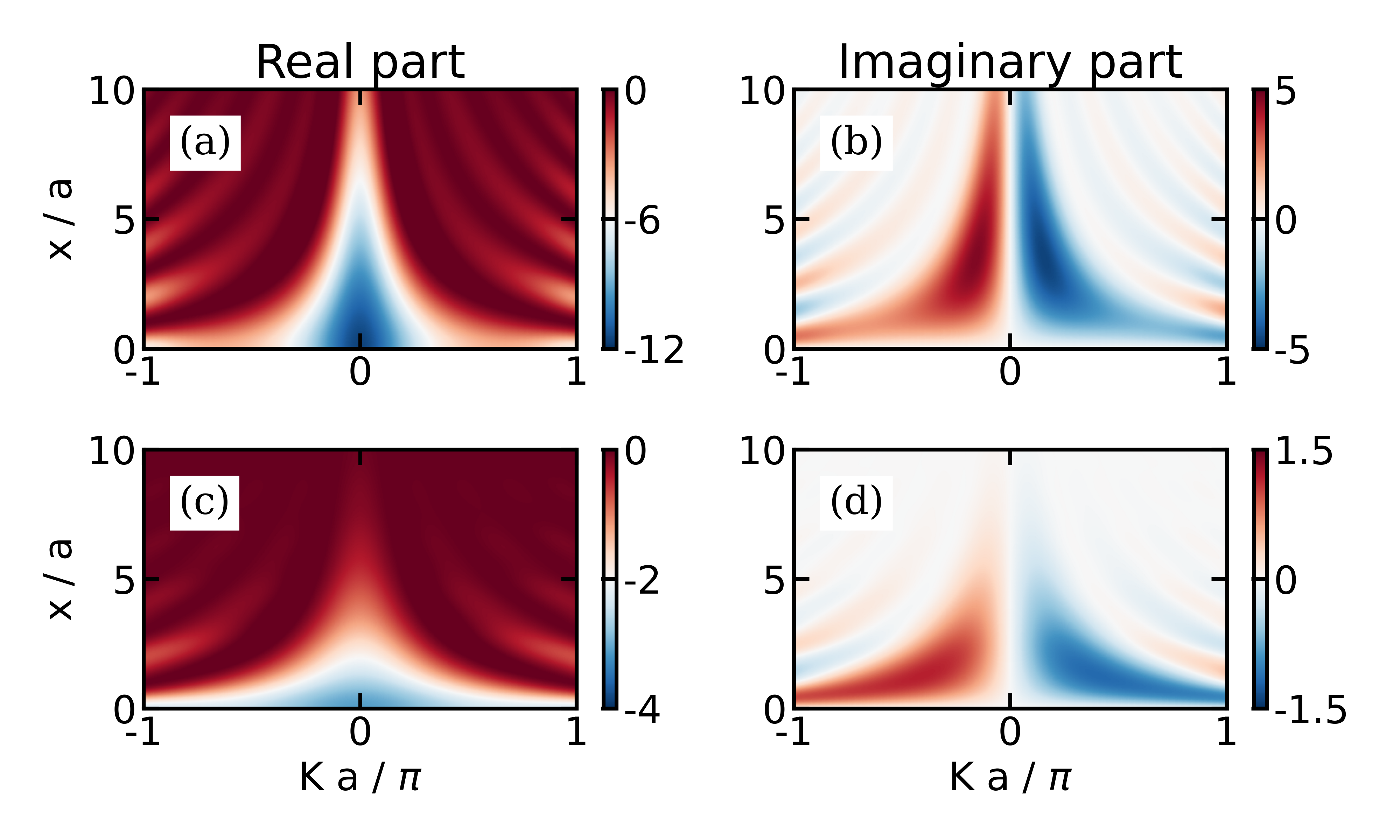}
\caption{Contour plots of the effective dimensionless interactions $\vec{F}_{K,b}(n_1,n_2)$ between the states $|e,e,\mbox{vac} \rangle$ and $\hat{P}_{K,b}^{\dagger}|g,g,\mbox{vac}\rangle$ as functions of $K a / \pi$ and $x/a$ for $\delta/J=0.0011$; to obtain the actual interaction strength, $\vec{F}_{K,b}(n_1,n_2)$ needs to be multiplied by $g^2 / (N^{1/2}J)$. (a) $\mbox{Re}[\vec{F}_{K,b}(n_1,n_2)]$ for $U/J=-1$. (b) $\mbox{Im}[\vec{F}_{K,b}(n_1,n_2)]$ for
$U/J=-1$. (c) $\mbox{Re}[\vec{F}_{K,b}(n_1,n_2)]$ for $U/J=-5/2$. (d) $\mbox{Im}[\vec{F}_{K,b}(n_1,n_2)]$ for $U/J=-5/2$. 
The color scheme for each of the four panels is different.
}
\label{fig_effective_fcapk}
\end{figure}

Figure~\ref{fig_effective_fcapk} shows
 $\vec{F}_{K,b}(n_1,n_2)$ as functions of $Ka/\pi$ and $x/a$
for $\delta/J=0.0011$ and two different $U/J$, namely
$U/J=-1$ (top row)
and $U/J=-5/2$ (bottom row).
The real and imaginary parts of
$\vec{F}_{K,b}(n_1,n_2)$ are shown in the left and right columns, respectively.
We note that
the system properties only depend
on the emitter separation $x/a$
and not independently 
on the actual emitter positions
$n_1$ and $n_2$; to make Fig.~\ref{fig_effective_fcapk}, the separation is---to aid with the visualization---treated as a continuous as opposed to a discrete variable.
The  magnitude of the real part of the 
effective interactions is larger for $U/J=-1$ (weakly-bound state) than for $U/J=-5/2$
(more strongly-bound state).
The characteristics common to both $U/J$ values
considered in Fig.~\ref{fig_effective_fcapk} are: 
First, the real part of the 
effective interactions is most negative near $Ka/\pi=x/a=0$, even though the resonance
wave vector $K^{(0)}$ differs in the two cases [$K^{(0)}a/\pi=0.0152$
for Figs.~\ref{fig_effective_fcapk}(a)/(b)
and $K^{(0)}a/\pi=0.0162$ for Figs.~\ref{fig_effective_fcapk}(c)/(d)].
Second, the real part of the effective interactions displays 
a larger $K$ dependence for $x/a=0$ than for $x/a>0$.
Third, for fixed $Ka/\pi$, 
the real part of the effective interactions is characterized by an overall fall-off that sits
on top of small amplitude oscillations.
Fourth, the magnitude of the imaginary part of the 
effective interactions is very small for $K a /\pi \approx 0$.
The separation and wave vector dependencies of $\vec{F}_{K,b}(n_1,n_2)$ have, as shown in the later sections, a strong impact on the system dynamics.

\begin{figure}
\includegraphics[width=0.48\textwidth]{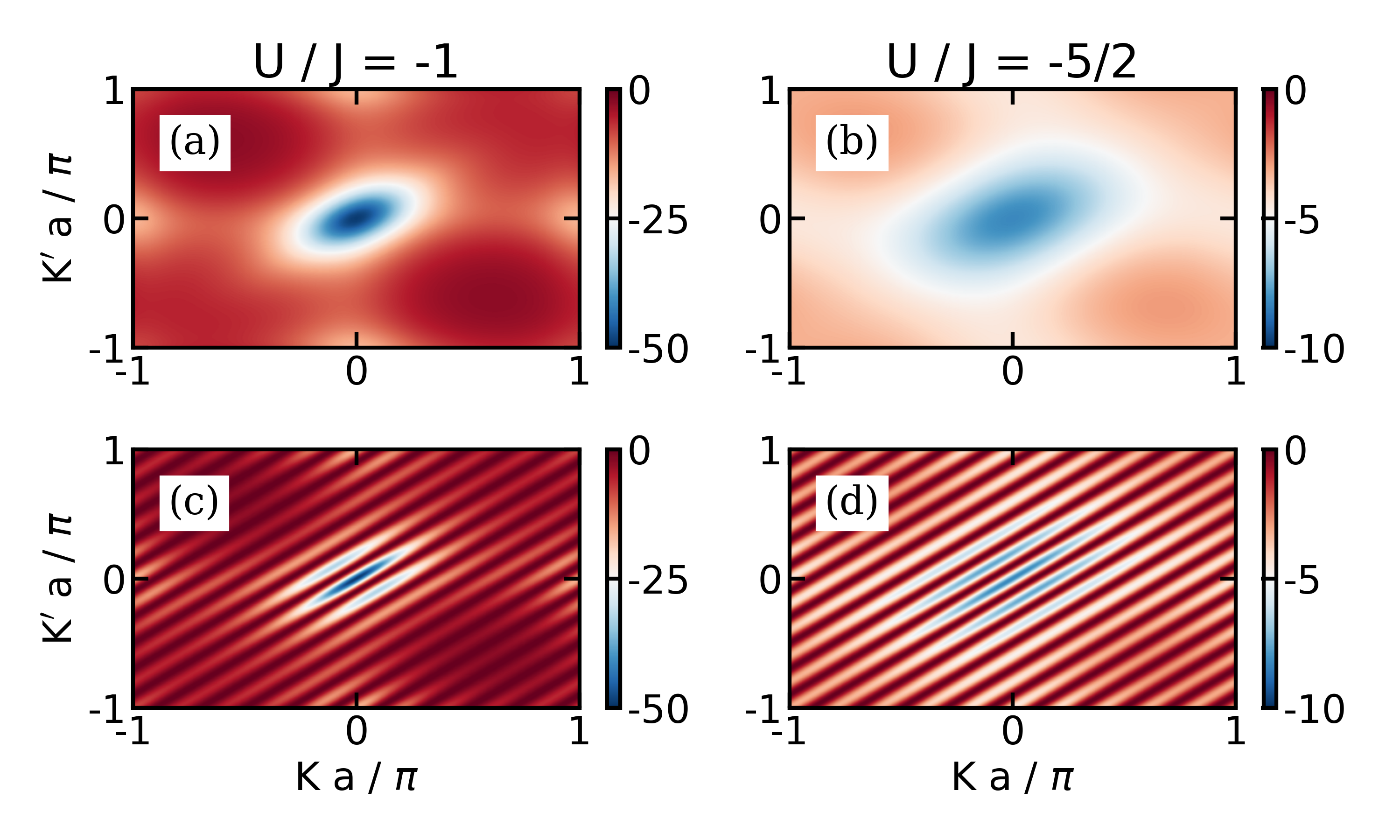}
\caption{Contour plots of the real part of the effective dimensionless  interactions
$\underline{G}_{K,K'}(n_1,n_2)$ between the states $|g,g,K\rangle$ and $|g,g,K'\rangle$ as functions of $Ka/ \pi$ and $K' a / \pi$ for $\delta/J=0.0011$; to obtain the actual interaction strength, $\underline{G}_{K,K'}(n_1,n_2)$ needs to be multiplied
by $g^2 / (NJ)$.
(a) $\text{Re}[\underline{G}_{K,K'}(n_1,n_2)]$
for $U/J=-1$ and $x/a=0$.
(b) $\text{Re}[\underline{G}_{K,K'}(n_1,n_2)]$
for $U/J=-5/2$ and $x/a=0$.
(c) $\text{Re}[\underline{G}_{K,K'}(n_1,n_2)]$
for $U/J=-1$ and $x/a=10$.
(d) $\text{Re}[\underline{G}_{K,K'}(n_1,n_2)]$
for $U/J=-5/2$ and $x/a=10$.
The color schemes for $U/J=-1$ [(a) and (c)] are the same;
similarly, the color schemes for $U/J=-5/2$ [(b) and (d)] are the same.
}
\label{fig_gkkprime}
\end{figure}

Section~\ref{sec_results} shows that the effective interactions $\underline{G}_{K,K'}(n_1,n_2)$
play a non-negligible role for scenarios (B) and (C), corresponding to the horizontal 
solid and dotted lines
in Fig.~\ref{fig1}(b).
The effective interactions $\underline{G}_{K,K'}(n_1,n_2)$ between two two-photon bound states, one with $K$ and the other 
with $K'$,
depend---for fixed $U/J$ and $\delta/J$---on $K a$, $K' a$, and $x/a$.
Figure~\ref{fig_gkkprime} shows the real part of
$\underline{G}_{K,K'}(n_1,n_2)$  for $\delta/J=0.0011$ for two different
separations, namely,
 $x=0$ (top row)
and $x/a=10$ (bottom row).
The left and right columns are for $U/J=-1$ and $U/J=-5/2$, respectively.
The key characteristics are:
(i) The oscillatory structure of the real 
part of $\underline{G}_{K,K'}(n_1,n_2)$
increases with increasing separation.
(ii) 
For the onsite interaction and detuning considered,
the real part of $\underline{G}_{K,K'}(n_1,n_2)$
is negative; the most negative values
are found for $K=K'=0$ for $x/a=0$ and $x/a=10$.
We note that the imaginary part of 
$\underline{G}_{K,K'}(n_1,n_2)$ (not shown) is zero for 
$K=K'$.

Since $\hat{H}^{\text{adia}}$ is hermitian [this can be seen readily by inspecting
$\underline{G}_{K,K'}(n_1,n_2)$], the population is normalized at all times, i.e.,
$|c_{ee}(t)|^2 + \sum_{K} |c_{K,b}(t)|^2=1$, and the eigen energies of $\hat{H}^{\text{adia}}$ are real.
The validity of the approximations 
(adiabatic elimination and dropping of
scattering states)
can thus be assessed in two complementary ways, namely by
 comparing the time evolution of, e.g., the initial state
$|e,e,\mbox{vac} \rangle$ under $\hat{H}$ and $\hat{H}^{\text{adia}}$ and by
comparing the eigen spectra of $\hat{H}$ and $\hat{H}^{\text{adia}}$.
Reference~\cite{ref_rabl_non-linear} constructed a master equation, using Eq.~(\ref{eq_hamiltonian_adia}) with
$G_{K_l,K_{l'}}(n_1,n_2)=0$ as a starting point. We denote $\hat{H}^{\text{adia}}$ with
$G_{K_l,K_{l'}}(n_1,n_2)=0$ and $\ne 0$ by $\hat{H}^{\text{adia},0}$ and  $\hat{H}^{\text{adia},1}$,
respectively.
Section~\ref{sec_results} shows that $\hat{H}^{\text{adia},1}$ significantly expands the applicability regime of the
reduced dimensionality Hamiltonian compared to $\hat{H}^{\text{adia},0}$
in certain parameter regimes.

\subsection{Markov approximation for $\hat{H}^{\text{adia},0}$}
\label{sec_markov_approximation}

For scenario (A), the population of the initial state $|e,e,\text{vac} \rangle$ decays approximately
exponentially for $x/a$ not too large. 
As shown in Ref.~\cite{ref_rabl_non-linear}, the decay constant can be determined analytically
in this regime using the Markov approximation.
Appendix~\ref{appendixA} shows that $\tilde{c}_{ee}(t)$,
where $\tilde{c}_{ee}(t)$ denotes the expansion coefficient for the state that rotates with $2 \Delta_e$, falls off exponentially according to
\begin{eqnarray}
\label{eq_exponential}
\tilde{c}_{ee}(t) = \exp \left( - \Gamma_{\text{bath}} t \right),
\end{eqnarray}
where $\Gamma_{\text{bath}}$ is given by
\begin{eqnarray}
\label{eq_gammabath}
\Gamma_{\text{bath}} = \frac{g^4 a}{ J^3 \hbar} |F_{K^{(*)},b}(n_1,n_2)|^2 \rho(K^{(*)}).
\end{eqnarray}
Here, $K^{(*)}$ is defined through 
\begin{eqnarray}
E_{K^{(*)},b} = 2 \hbar \omega_e +2 \Delta_e,
\end{eqnarray}
with the ``Stark shift" $2\Delta_e$~\cite{ref_rabl_non-linear},
\begin{eqnarray}
\label{eq_appendixA_last}
2\Delta_e = -\frac{2}{N} \sum_k \frac{g^2}{\Delta_k },
\end{eqnarray}
quantifying the shift of the state $|e,e,\mbox{vac} \rangle$
due to the ``renormalization" by the single-photon states. 
Correspondingly, the decoupled ($g=0$) resonance wave number
$K^{(0)}$ gets shifted to $K^{(*)}$ for finite $g/J$; 
the use of $K^{(*)}$ in place of $K^{(0)}$, as done in
Ref.~\cite{ref_rabl_non-linear}, provides an improved description.
The density of states $\rho(K^{(*)})$ at the resonance wave vector can be written as
\begin{eqnarray}
\rho(K^{(*)}) = J \left(\hbar v_{K^{(*)},b} \right)^{-1}.
\end{eqnarray}
When $|g/J|$ is not
much smaller than $1$, the 
adiabatic elimination and, correspondingly, the concept of a resonant wave number looses its meaning.
The importance of the Stark shift $2\Delta_e$ increases as  $K^{(0)}a$ and, correspondingly,
the detuning $\delta/J$ approach zero.

\begin{figure}
\includegraphics[width=0.47\textwidth]{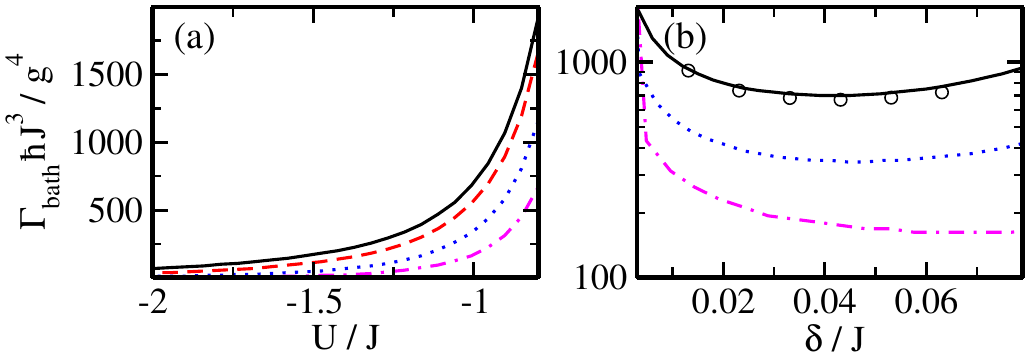}
\caption{Lines show the dimensionless decay constant $\Gamma_{\text{bath}} \hbar J^3 /g^4$, obtained within the Markov approximation [Eq.~(\ref{eq_gammabath})].
(a)
The black solid, red dashed, blue dotted, and magenta dash-dotted lines
show $\Gamma_{\text{bath}} \hbar J^3 /g^4$ 
as
a function of $U/J$ for $\delta/J=0.0431$ 
and
$x/a=0$, $2$, $4$, and $6$, respectively. 
(b)
The black solid, blue dotted, and magenta dash-dotted lines
show $\Gamma_{\text{bath}} \hbar J^3 /g^4$ 
as
a function of $\delta/J$ for $U/J=-1$ and $x/a=0$, $4$, and $6$, respectively. 
For comparison, the open black circles show
the decay constant extracted from the dynamics for the full Hamiltonian $\hat{H}$ for $x/a=0$ and $g/J=1/50$; the Markov approximation results (solid line) capture the decay constant extracted from the full decay dynamics quite well. 
Note that the Markov approximation breaks down
when $\delta/J$ approaches zero (left portion
of the panel) and when $\delta/J$
approaches the two-photon scattering continuum
(right portion of the panel).
}
\label{fig_gamma}
\end{figure}

Figures~\ref{fig_gamma}(a) and \ref{fig_gamma}(b) show the decay constant $\Gamma_{\text{bath}}\hbar J^3 /g^4$
as a function of the onsite interaction $U/J$ and the detuning
$\delta/J$, respectively, for various separations
($x/a=0$ to $6$).
It can be seen that the radiation dynamics is characterized by a larger dimensionless 
decay constant (faster decay) for $x/a=0$ (solid line) than for $x/a=6$
(dash-dotted line).
This makes sense intuitively since a larger separation is
associated with a smaller, in magnitude, effective interaction $F_{K^{(*)},b}(n_1,n_2)$.
The dependence on $U/J$ [see Fig.~\ref{fig_gamma}(a)] can also be understood 
readily intuitively.
As $|U/J|$ increases, the two-photon bound state becomes
more localized and the coupling to the state $|e,e,\text{vac}\rangle$
decreases. The Markov approximation results shown in Fig.~\ref{fig_gamma}(a)
agree quite well with the decay constants extracted from full numerical calculations (not shown). 

The dependence of the dimensionless decay constant,
calculated within the Markov approximation, on the detuning 
is non-monotonic
[see Fig.~\ref{fig_gamma}(b)]. The increase of $\Gamma_{\text{bath}}\hbar J^3 /g^4$
as the dimensionless detuning $\delta/J$,
for fixed $x/a$,
approaches zero [left part of Fig.~\ref{fig_gamma}(b)]
is unphysical. This increase is due to the break-down of the Markov
approximation in the vicinity of the bottom of the band,
where the density of states of two-photon bound states is  large
and diverges as $\delta/J \rightarrow 0$. 
The open circles in Fig.~\ref{fig_gamma}(b) show the decay constant for $x/a=0$, extracted from calculations for the full Hamiltonian $\hat{H}$. While the agreement with the Markov approximation results is quite good, we note that the full dynamics displays non-exponential characteristics for small $\delta/J$ that get ``averaged" when fitting to an exponential.
The Markov approximation also breaks down when $\delta/J$ becomes too large 
[right part of Fig.~\ref{fig_gamma}(b)]. The reason for this break-down is that the adiabatic elimination is not valid when $2 \hbar \omega_e$ is close to the two-photon scattering continuum.

To recapitulate, we arrived at Eq.~(\ref{eq_exponential}) by making four distinct 
approximations: adiabatically eliminating $c_{1k}(t)$ and $c_{2k}(t)$, neglecting the two-photon scattering
continuum, neglecting $\underline{G}_{K,K'}(n_1,n_2)$, and making the Markov approximation.
It is useful to compare the results obtained for the two-emitter case with non-linear bath to those 
for a single emitter [$\hat{H}$ in Eq.~(\ref{eq_ham_total}) with 
 $N_e=1$ and $U=0$ with initial state $|e,\text{vac} \rangle$]. 
 Appendix~\ref{appendix_single_emitter} shows that
 the decay constant $\Gamma_{\text{single}}$ for the
 single-emitter system evaluates within the Markov 
 approximation to   
 \begin{eqnarray}
 \label{eq_gammasingle}
 \Gamma_{\text{single}}= \frac{g^2 a}{\hbar J}\rho_{\text{single}}(k^{(0)}),
 \end{eqnarray}
 where 
 \begin{eqnarray}
\rho_{\text{single}}(k^{(0)}) = J \left(\hbar v_{k^{(0)}} \right)^{-1},
\end{eqnarray}
with $k^{(0)}$ denoting the single-photon resonance wave vector.

Comparison of Eqs.~(\ref{eq_gammabath}) and (\ref{eq_gammasingle})
indicates that the two-emitter dynamics, in the regime where $|c_{ee}(t)|^2$---starting in the state
$|e,e,\text{vac} \rangle$---falls off exponentially, is the same as that for the single emitter system, 
provided (i) the dimensionless densities of states
$a \rho(K^{(*)})$ and $a \rho_{\text{single}}(k^{(0)})$ take the same value and 
(ii) the coupling constant $g_{\text{single}}$ of the single emitter system
is set to 
\begin{eqnarray}
\label{eq_geffective_single_emitter}
g_{\text{single}} = \frac{g^2}{ J } |F_{K^{(*)},b}(n_1,n_2)|;
\end{eqnarray}
the quantities on the right hand side are understood to be those characterizing the two-emitter system.
To match the densities of states, we consider $K^{(*)}a$ and $k^{(0)}a$
values that are sufficiently large for the Markov approximation to hold but sufficiently small for
$\Delta_{K,b}$ and $E_k$ to be well approximated by their Taylor-expanded expressions up to order 
$(Ka)^2$ and $(ka)^2$, respectively.
Comparing the slopes of the quadratic terms, we find that the dimensionless densities of states match if the
tunneling coupling strength
$J_{\text{single}}$
of the single emitter system is set to
\begin{eqnarray}
\label{eq_jeffective_single_emitter}
J_{\text{single}} = 2 J\left[ \left( \frac{U}{J} \right)^2 + 16 \cos^2 \left( \frac{K^{(*)}a}{2} \right) \right]^{-1/2} ;
\end{eqnarray}
the quantities on the right hand side are, again, understood to be those characterizing the two-emitter system.

The meaning of Eqs.~(\ref{eq_geffective_single_emitter}) and (\ref{eq_jeffective_single_emitter}) is as follows. Say we have a coupled two-emitter--cavity system in the Markovian regime. For a given $U/J$, $g/J$, $\delta/J$, and $x/a$, this implies that the exponential decay of the population
is characterized by $\Gamma_{\text{bath}}$.
Imagine now that we want to design a coupled single-emitter--cavity system 
such that the exponential decay
of $|d_e(t)|^2$, see Eq.~(\ref{eq_appendixA_first}), is characterized by $\Gamma_{\text{single}}=\Gamma_{\text{bath}}$.
This goal is accomplished if the tunneling coupling strength $J_{\text{single}}$ and coupling strength $g_{\text{single}}$ of the 
single-emitter--cavity system are chosen according to
Eqs.~(\ref{eq_geffective_single_emitter}) and (\ref{eq_jeffective_single_emitter}).

Using that $|F_{K^{(*)},b}(n_1,n_2)|$ is---for the parameters considered in this paper 
(see Fig.~\ref{fig_effective_fcapk} for two examples)---of the order of $1$ to $10$, Eq.~(\ref{eq_geffective_single_emitter}) shows that the 
two-emitter dynamics considered is slower than the single-emitter dynamics would be
if the single emitter was in resonance with the single-photon band.
Importantly, if the
two-emitter energy is in resonance with the two-photon bound state band, then the single-emitter energy is not in resonance with the single-photon band
(at least not for the parameters considered in this work). 
We note that appreciable single-emitter dynamics is observed for large $x/a$
for certain parameter combinations (see Sec.~\ref{sec_results} for details).

\section{Stationary solution}
\label{sec_results_static}

This section discusses the stationary solutions of the coupled emitter-cavity system under study.
Section~\ref{sec_overview}
provides an overview for different detunings while Sec.~\ref{sec_rabi} focuses on the physics near the bottom of the band.

\subsection{Overview}
\label{sec_overview}

\begin{figure}
\includegraphics[width=0.3\textwidth]{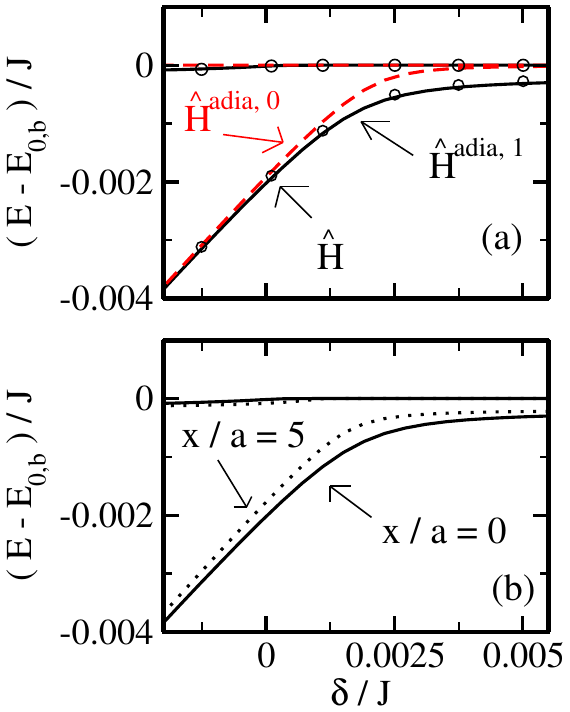}
\caption{ 
Energy of the two lowest eigen  states as a function of $\delta/J$ for $U/J=-1$ and $g/J=1/50$.
(a) The black solid and red dashed lines show 
the energy for
$\hat{H}^{\text{adia},1}$ and 
$\hat{H}^{\text{adia},0}$, respectively, for $x=0$. For comparison, the open circles show the eigen energies for $\hat{H}$ (using a basis that excludes the two-photon scattering states).
The energies for $\hat{H}$
and $\hat{H}^{\text{adia},1}$ agree very well.
(b) The black solid and black dotted lines show the energies of $\hat{H}^{\text{adia},1}$  for $x/a=0$  and $x/a=5$, respectively. A clear separation dependence can be seen.
In both panels, the lower state corresponds to a hybridized bound state
with appreciable $|e,e,\text{vac} \rangle$ and $|g,g,\text{pol} \rangle$ contributions
(see Sec.~\ref{sec_rabi} for details).
}
\label{fig_energy}
\end{figure}

To assess the reliability of the different approximations, we compare the energy spectrum obtained by diagonalizing
$\hat{H}^{\text{adia,0}}$, 
$\hat{H}^{\text{adia,1}}$, and 
$\hat{H}$ (using a basis that excludes the two-photon scattering states) for $U/J=-1$ for various $\delta/J$.
Figure~\ref{fig_energy}(a) shows the lowest two eigen energies for
$x/a=0$.
The zero of the energy axis corresponds to the bottom of the $g=0$ two-photon energy band.
The eigen
energies of the full Hamiltonian (open circles)
are reproduced much better by 
$\hat{H}^{\text{adia,1}}$ (black solid lines)
than by
$\hat{H}^{\text{adia,0}}$ (red dashed lines).
Specifically, neglecting the 
effective interactions $\underline{G}_{K,K'}(n_1,n_2)$
leads to a weakening of the binding of the lowest energy state, especially for positive $\delta/J$.
The lowest energy state is a hybridized bound state that contains contributions from the state $|e,e,\text{vac}\rangle$ and two-photon bound states. The hybridized state is clearly separated from the energy continuum. The character of the lowest energy eigen state 
is elucidated in the next section.

\begin{figure}
\includegraphics[width=0.44\textwidth]{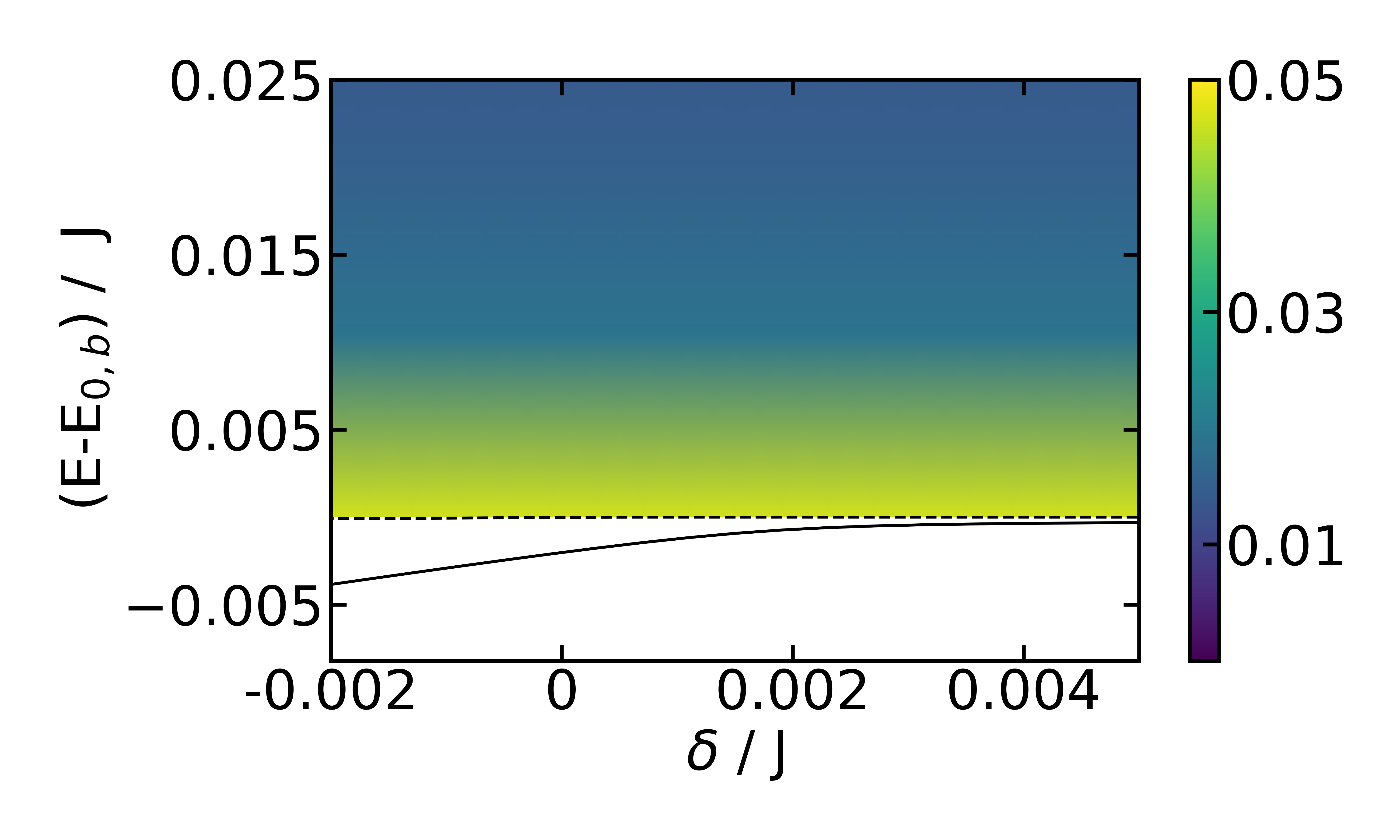}
\caption{Visualization of eigen spectrum 
of $\hat{H}^{\text{adia},1}$
as a function of the detuning $\delta / J$ for $U/J=-1$, $g/J=1/50$, and $x=0$.
The density of states $\rho_E(E)$ of the
continuum 
portion of the energy spectrum, which is dominated by states that 
have no or extremely small emitter admixtures,
is shown in color (the legend is shown on the right; arbitrary units are used). 
The lowest hybridized bound state (black solid line) is well separated from
the energy continuum.
The second lowest state (black dashed line) is separated by a small gap from the continuum for negative $\delta/J$ 
and part of the continuum for positive $\delta/J$.
}
\label{fig_DOS}
\end{figure}

The density of states $\rho_E(E)$ (number of states per unit energy interval) of the energy continuum,
which reduces to the two-photon bound state band
for $g=0$, is shown by the color map in Fig.~\ref{fig_DOS}
for the same parameters as those used in Fig.~\ref{fig_energy}(a).
The density of states is large near the bottom of the energy band and decreases as one moves away from the bottom of the band.
While the coupling constant $g$ has a profound effect on the two lowest eigen states (black solid and dashed lines in Fig.~\ref{fig_DOS}), the density of states depends comparatively weakly on $g$.

To elucidate the dependence on the separation for $U/J=-1$,
we work with $\hat{H}^{\text{adia,1}}$. 
Figure~\ref{fig_energy}(b) shows the
two lowest eigen energies as a function of $\delta/J$ for $x/a=0$ (solid line) and $x/a=5$ (dotted line).
The binding energy of the lowest hybridized bound state decreases with increasing $x/a$.
This
might be expected naively since a larger emitter separation
is associated with reduced interactions.
Interestingly, the second lowest energy state has a somewhat lower 
energy for $x/a=5$ than for $x/a=0$; this can be seen 
most clearly in Fig.~\ref{fig_energy}(b) for negative detuning
but also holds true for positive $\delta/J$.
An analysis of the corresponding eigen state
reveals that the second lowest state for $x/a=5$ has bound state character not only for negative but also for positive detuning. For $x/a=0$, in contrast, the 
second lowest state merges into the
continuum when the detuning is positive. 
The emergence of a second bound state with increasing 
separation is somewhat counterintuitive.
We checked that the full Hamiltonian $\hat{H}$ 
also supports a second bound state, i.e., we checked that its appearance 
is not an artefact of the adiabatic approximation.
To gain additional insights, the next section discusses a
two-state model that captures key aspects of the 
two hybridized bound states that exist for $U/J=-1$, small detuning, and sufficiently large $x/a$.

 \subsection{Near the bottom of the band: Two-state model}
 \label{sec_rabi}
 
 An important conclusion of the previous section is that the adiabatic Hamiltonian $\hat{H}^{\text{adia,1}}$ captures the key features of the bound states supported by the full Hamiltonian $\hat{H}$. Using $\hat{H}^{\text{adia,1}}$, we now review the  physical picture  that was introduced in Ref.~\cite{ref_jugal}. Since the effective interactions $\underline{G}_{K,K'}(n_1,n_2)$ cannot be neglected near the bottom of the band (see the previous section), the bath Hamiltonian contains off-diagonals in the $\{|g,g,K\rangle,|e,e,\text{vac}\rangle \}$ basis. To proceed, we change the basis. We continue to use $|g,g \rangle$  and $|e,e\rangle$ with energy $0$ and $2 \Delta_e$, respectively, for the two emitters.
 For the two-photon bath Hamiltonian, in contrast, we change from the basis   states $|K \rangle$, in which the bath is characterized by effective interactions between two-photon bound states with center-of-mass wave numbers $K$ and $K'$, 
 \begin{eqnarray}
 \underline{H}_{\text{b}}^{\text{adia,1}}=
 \underline{\Delta}_{K,b} + \frac{g^2}{JN} \underline{G}_{K,K'}(n_1,n_2),
 \end{eqnarray}
 to a basis in which the bath Hamiltonian
 $\hat{H}_{\text{b}}^{\text{adia,1}}$
 is diagonal. 
 Performing the diagonalization,
we find that the energy spectrum of the adiabatic bath Hamiltonian $\hat{H}_{\text{b}}^{\text{adia,1}}$
 consists of a continuum, similar to the two-photon bound state band, and an ``isolated state" whose energy is energetically separated from the bottom of the two-excitation continuum; this state lives in the band gap and corresponds to a bound state.
 
 The 
isolated state $|\text{pol} \rangle$ is well reproduced by an ansatz with Lorentzian distributed expansion coefficients $d_K$~\cite{ref_jugal},
\begin{eqnarray}
\label{eq_wavefunc-pol}
|\text{pol}\rangle = \sum_K d_K |K \rangle
\end{eqnarray}
and 
\begin{eqnarray}
d_K = \frac{(2L_{\text{eff}}^{-1}a)^{3/2}}{\sqrt{N}}
\frac{1}{(2Ka)^2 +(L_{\text{eff}}^{-1}a)^2},
\end{eqnarray} 
where the normalization 
is chosen such that
\begin{eqnarray}
 a \int_{-\infty}^{\infty} |d_K|^2 dK=1.
 \end{eqnarray}
In writing this ansatz, it is assumed that $L_{\text{eff}}^{-1}a$ is much smaller than $\pi$ so that the integration limits can be safely extended 
from $\pm \pi$ to $\pm \infty$. 
We refer to the isolated state as a polaron-like state as it represents a quasi-particle that is a superposition of states with different center-of-mass momenta.
The wave number width of the expansion coefficients $d_K$ is given by $(L_{\text{eff}})^{-1}$.
We determine $L_{\text{eff}}$  by minimizing the ground state energy of $\hat{H}^{\text{adia,1}}_{\text{b}}$.
To make the calculations tractable analytically, we approximate 
the effective interactions $\underline{G}_{K,K'}(n_1,n_2)$ by a
constant, namely their value at 
$K=K'=K^{(0)}$.
We find
\begin{eqnarray}
\frac{L_{\text{eff}}}{a}=\frac{2J^3}{g^2|G_{K^{(0)},K^{(0)}}(n_1,n_2)|\sqrt{U^2+16J^2}}
\end{eqnarray}
and
\begin{eqnarray}
\label{eq_e-pol}
E_{\text{pol}} = -\delta-\frac{g^4}{8J^4}|G_{K^{(0)},K^{(0)}}(n_1,n_2)|^2 \sqrt{U^2+16J^2}.
\end{eqnarray} 
 This variational result
 reproduces the numerically determined ground state energy
 of $\hat{H}^{\text{adia,1}}_{\text{b}}$ very
 well. 
 In the condensed matter context, the Hamiltonian that supports the polaron-like state shows up when an impurity or defect in a one-dimensional lattice is associated with attractive all-to-all momentum space interactions.
 All-to-all  interactions are currently being investigated  by a number of groups due to their relevance in quantum gravity and spin glass physics~\cite{ref_periwal,ref_gadway,ref_rey}.
 
\begin{figure}
\includegraphics[width=0.48\textwidth]{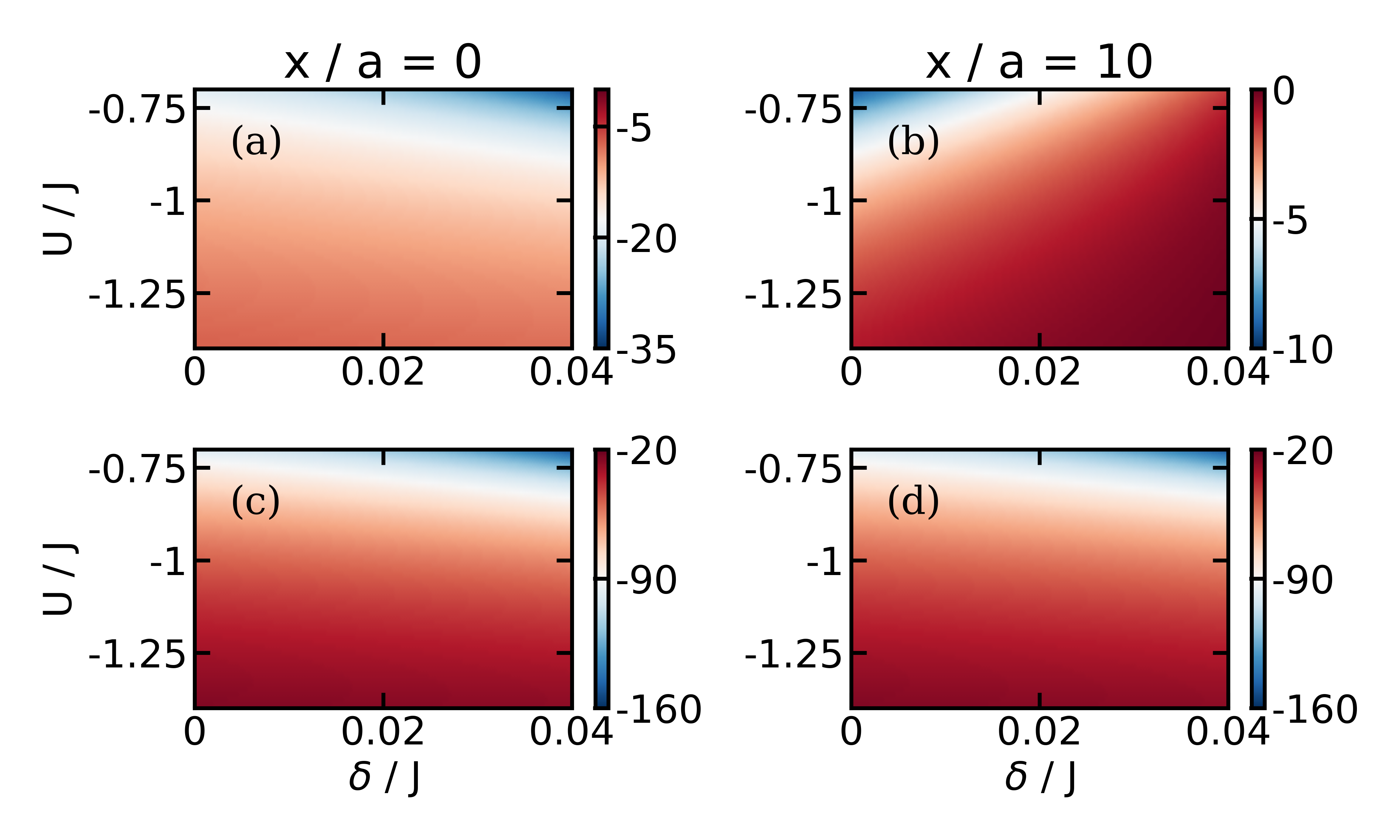}
\caption{Contour plots of the effective dimensionless interaction ${F}_{K^{(0)},b}(n_1,n_2)$ between the states $|e,e,\mbox{vac} \rangle$ and $\hat{P}_{K^{(0)},b}^{\dagger}|g,g,\mbox{vac}\rangle$ and the effective dimensionless interaction ${G}_{K^{(0)},K^{(0)}}(n_1,n_2)$ between the states $|g,g,K^{(0)}\rangle$ and $|g,g,K^{(0)}\rangle$ as functions of $\delta/J$ and $U/J$; 
to obtain the actual interaction strengths, ${F}_{K^{(0)},b}(n_1,n_2)$ and ${G}_{K^{(0)},K^{(0)}}(n_1,n_2)$ need to be multiplied by $g^2 / (N^{1/2}J)$ and $g^2 / (NJ)$, respectively. (a) Re$[{F}_{K^{(0)},b}(n_1,n_2)]$ for $x/a=0$. (b) Re$[{F}_{K^{(0)},b}(n_1,n_2)]$ for $x/a=10$. (c) Re$[{G}_{K^{(0)},K^{(0)}}(n_1,n_2)]$ for $x/a=0$. (d) Re$[{G}_{K^{(0)},K^{(0)}}(n_1,n_2)]$ for $x/a=10$. The color schemes for Re$[{F}_{K^{(0)},b}(n_1,n_2)]$ are different for the two separations. The color schemes for Re$[{G}_{K^{(0)},K^{(0)}}(n_1,n_2)]$,
in contrast, are the same for the two separations.}
\label{fig_GKK_fK}
\end{figure}

 As already alluded to in Sec.~\ref{sec_adiabatic_elimination},
the effective interactions
 $\underline{G}_{K,K'}(n_1,n_2)$ are essentially purely real for $K=K'$.
 Figures~\ref{fig_GKK_fK}(c) and \ref{fig_GKK_fK}(d) show Re$[G_{K^{(0)},K^{(0)}}(n_1,n_2)]$ as functions of $\delta/J$ and $U/J$ for $x/a=0$ and $x/a=10$, respectively. 
 It can be seen that $G_{K^{(0)},K^{(0)}}(n_1,n_2)$ depends extremely weakly on the separation $x/a$. Consequently, the energy $E_{\text{pol}}$ of the photonic polaron is to a very good approximation independent of the emitter separation $x/a$.

 Next, we rewrite $\hat{H}^{\text{adia,1}}$ in the 
 product basis in which the emitter and bath Hamiltonians are
 diagonal.
 Transforming the system-bath coupling 
 $g^2 N^{-1/2} \vec{F}_{K,b}(n_1,n_2)/J$
 to the new basis and restricting the Hilbert space to the states
 $|e,e,\text{vac}\rangle$ and $|g,g,\text{pol}\rangle$, we arrive at the following matrix
 representation of the two-state Hamiltonian $\hat{H}^{\text{2-st.}}$:
 \begin{eqnarray}
 \label{eq_ham_two_state}
  \underline{H}^{\text{2-st.}}=
  \left(
  \begin{array}{cc}
  2 \Delta_e & G_{\text{eff}}(n_1,n_2) \\
  \left[G_{\text{eff}}(n_1,n_2)\right]^* & E_{\text{pol}}
  \end{array}\right).
 \end{eqnarray}
Using our variational expression for $|g,g,\text{pol} \rangle$, the 
effective coupling $G_{\text{eff}}(n_1,n_2)$ between states $|e,e,\text{vac}\rangle$ and $|g,g,\text{pol} \rangle$ can be written as
\begin{eqnarray}
\label{eq_geff}
G_{\text{eff}}(n_1,n_2) 
= 
\frac{g^3(U^2+16J^2)^{1/4}}{2J^{5/2}} \nonumber \\
 \times F_{K^{(0)},b}(n_1,n_2)|G_{K^{(0)},K^{(0)}}(n_1,n_2)|^{1/2}.
\end{eqnarray} 
 Figures~\ref{fig_GKK_fK}(a) and \ref{fig_GKK_fK}(b) show 
 Re$[{F}_{K^{(0)},b}(n_1,n_2)]$ as functions of $\delta/J$ and $U/J$ for $x/a=0$ and $x/a=10$, respectively. It can be seen that Re$[F_{K^{(0)},b}(n_1,n_2)]$  
 [Figs.~\ref{fig_GKK_fK}(a)-\ref{fig_GKK_fK}(b)]
 varies much more strongly with $x/a$ than $\text{Re}[G_{K^{(0)},K^{(0)}}(n_1,n_2)]$
 [Figs.~\ref{fig_GKK_fK}(c)-\ref{fig_GKK_fK}(d)]. We conclude that,
 in the regime where the two-state model 
 $\hat{H}^{\text{2-st.}}$ provides a faithful description, the separation dependence of the
 hybridized energy eigen states of $\hat{H}$ 
 (see the discussion surrounding Figs.~\ref{fig_energy} and \ref{fig_DOS}) 
 is due to the dependence of $G_{\text{eff}}(n_1,n_2)$
 on $F_{K^{(0)},b}(n_1,n_2)$.
 
 The eigen states $\Psi_{\pm}$ and eigen energies $E_{\pm}$ of $\hat{H}^{\text{2-st.}}$ read
\begin{eqnarray}
\Psi_{\pm}= d^{(\pm)}_{\text{vac}} |e,e,\text{vac} \rangle + d^{(\pm)}_{\text{pol}} |g,g,\text{pol} \rangle 
\end{eqnarray}
and
\begin{eqnarray}
E_{\pm}= \Big(\Delta_e +\frac{E_{\text{pol}}}{2}\Big) \pm
\sqrt{\Big(\Delta_e -\frac{E_{\text{pol}}}{2}\Big)^2 + |G_{\text{eff}}|^2}, 
\end{eqnarray}
where 
the expansion coefficients
$d^{(\pm)}_{\text{vac}}$ and
$d^{(\pm)}_{\text{pol}}$ are given by
\begin{eqnarray}
\label{eq_coeff1}
d^{(\pm)}_{\text{vac}}= N_{\pm} G_{\text{eff}} 
\end{eqnarray}
and
\begin{eqnarray}
\label{eq_coeff2}
\nonumber d^{(\pm)}_{\text{pol}}=  \\N_{\pm} \left(-\Delta_e +\frac{E_{\text{pol}}}{2} \pm
\sqrt{\left(\Delta_e -\frac{E_{\text{pol}}}{2}\right)^2 + |G_{\text{eff}}|^2}\right),
\end{eqnarray}
respectively; in Eqs.~(\ref{eq_coeff1})-(\ref{eq_coeff2}),
$N_{+}$ and $N_-$ denote normalization constants.
We refer to $\Psi_+$ and $\Psi_-$ 
as symmetric hybridized state and anti-symmetric hybridized state, respectively.

\begin{figure}
\includegraphics[width=0.3\textwidth]{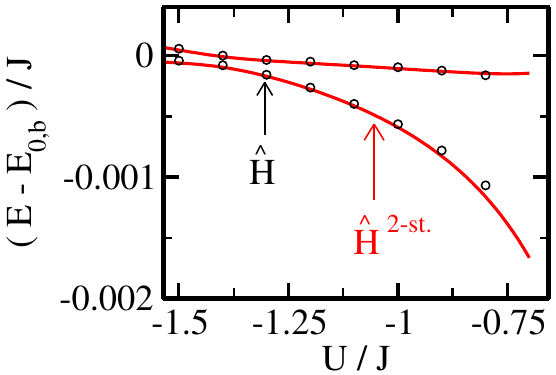}
\caption{Energy of the hybridized states as a function 
of $U/J$ for $g/J=1/50$, $\delta/J=0.0011$, and $x/a=10$.
The black circles and red lines are obtained using the 
full Hamiltonian $\hat{H}$ and the two-state 
Hamiltonian $\hat{H}^{\text{2-st.}}$, respectively.
The agreement is very good for the parameter regime considered.}
\label{fig_2-st}
\end{figure}

 Figure~\ref{fig_2-st} compares the two eigen energies supported by $\hat{H}^{\text{2-st.}}$ (red solid lines) for $\delta/J=0.0011$ and $x/a=10$ with the two eigen energies 
 of $\hat{H}$, whose eigen states have the largest overlap with the initial state $|e,e,\text{vac} \rangle$ (black circles), as a function of $U/J$.
 The two-state model reproduces the energy of the hybridized energy eigen
 states of the full Hamiltonian well. 
The state $\Psi_-$ is bound for $|U/J|$ values smaller than $1.4$ and unbound for $|U/J|$ values larger 
than $1.4$. Since a strong onsite interaction (large $|U/J|$) corresponds to more localized two-photon bound states (in real space), the ``reach" of the two-photon bound state for large $|U/J|$ is too small to induce
a new bound state.
As discussed in the next section, 
the two-state Hamiltonian 
$\hat{H}^{\text{2-st.}}$ describes several key characteristics of the 
dynamics predicted by 
the full Hamiltonian in the $|\delta/J| \rightarrow 0$ limit.

\section{Dynamics}
\label{sec_results}
\begin{figure}
\includegraphics[width=0.43\textwidth]{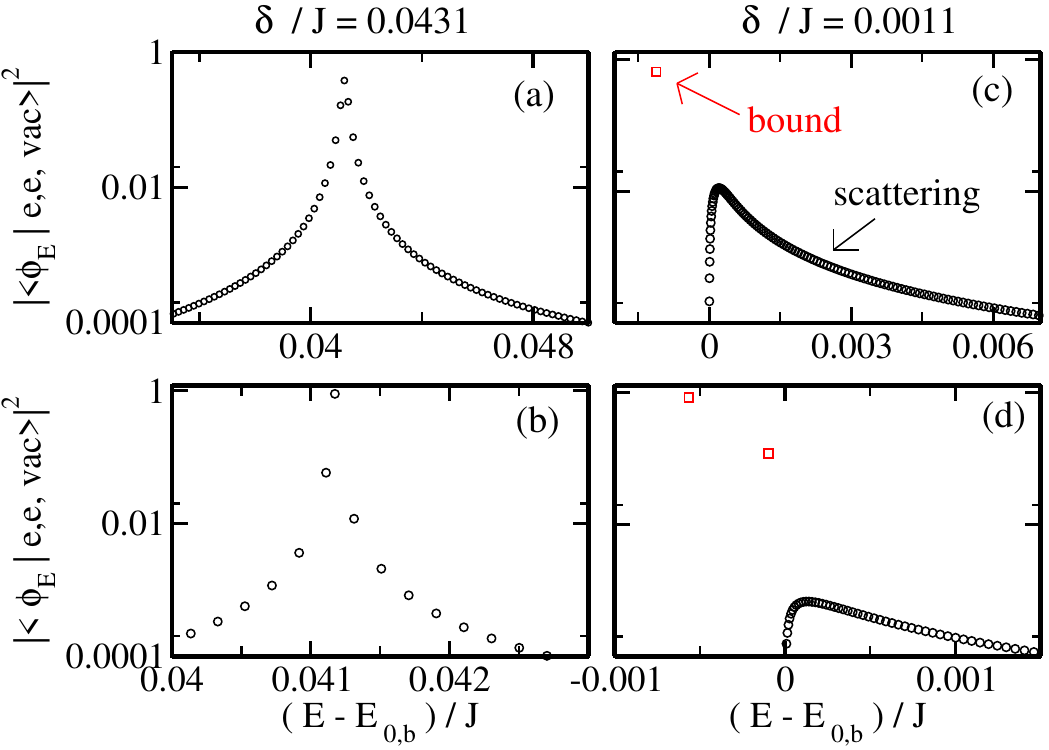}
\caption{Projection of the initial state $|e,e,\text{vac} \rangle$ onto the energy eigen states $\phi_E$ of $\hat{H}$ for $U/J=-1$ and $g/J=1/50$. The square of the absolute value of the overlap onto scattering states and bound states is shown by black circles and red squares, respectively, for
(a) $\delta/J=0.0431$ and $x/a=0$,
(b) $\delta/J=0.0431$ and $x/a=10$,
(c) $\delta/J=0.0011$ and $x/a=0$,
and
(d) $\delta/J=0.0011$ and $x/a=10$.}
\label{fig_overlap}
\end{figure}

This section discusses the radiation dynamics for $U/J=-1$ for various detunings $\delta/J$  and separations $x/a$.
Throughout, the 
initial state is taken to be the excited emitter state $|e,e,\text{vac} \rangle$.
Figure~\ref{fig_overlap} shows the decomposition of the state $|e,e,\text{vac} \rangle$
into the energy eigen states $\phi_E$ of $\hat{H}$ for
$U/J=-1$ and two different separations, namely, $x/a=0$ (top row) and $x/a=10$ (bottom row).
Two different detunings are considered:
$\delta/J=0.0431$ (left column) and $\delta/J=0.0011$ (right column).
As discussed in Sec.~\ref{sec_markov_approximation}, the Markov approximation provides a good description of the radiation dynamics for $\delta/J=0.0431$  but breaks down for 
$\delta/J=0.0011$.
For the larger detuning, the initial state is dominated by a few eigen states
whose energy is close to those corresponding to $K^{(*)}$.
The applicability of the Markov approximation relies on the fact that the overlap coefficients peak around one energy value and fall off quickly away from this energy. We emphasize that the overall behavior of the overlap coefficients for $x/a=0$ [Fig.~\ref{fig_overlap}(a)] and
$x/a=10$ [Fig.~\ref{fig_overlap}(b)]
is similar. Note, however, that the scale of the energy axis and the number of eigen states that contribute are significantly smaller for $x/a=10$ than for $x/a=0$.

As the detuning decreases to  small positive values, where the two-emitter energy
is very close to the bottom of the energy band, the decomposition of the initial state into the energy eigen states changes significantly.
For $x/a=0$, the initial state is dominated by a single state [red square in Fig.~\ref{fig_overlap}(c)] whose energy is separated from the energy continuum (round circles). Altogether, the states corresponding to the energy continuum contribute 34.5~\%.
For $x/a=10$, in contrast, there are two energy eigen states that contribute appreciably [84.6~\% and 12.0~\%; see red squares in
Fig.~\ref{fig_overlap}(d)]. 
 The eigen states corresponding to the red squares
 in Figs.~\ref{fig_overlap}(c) and \ref{fig_overlap}(d) are quite well described by the two-state model Hamiltonian $\hat{H}^{\text{2-st.}}$,
 Eq.~(\ref{eq_ham_two_state}).

\begin{widetext}

\begin{figure}
\includegraphics[width=0.7\textwidth]{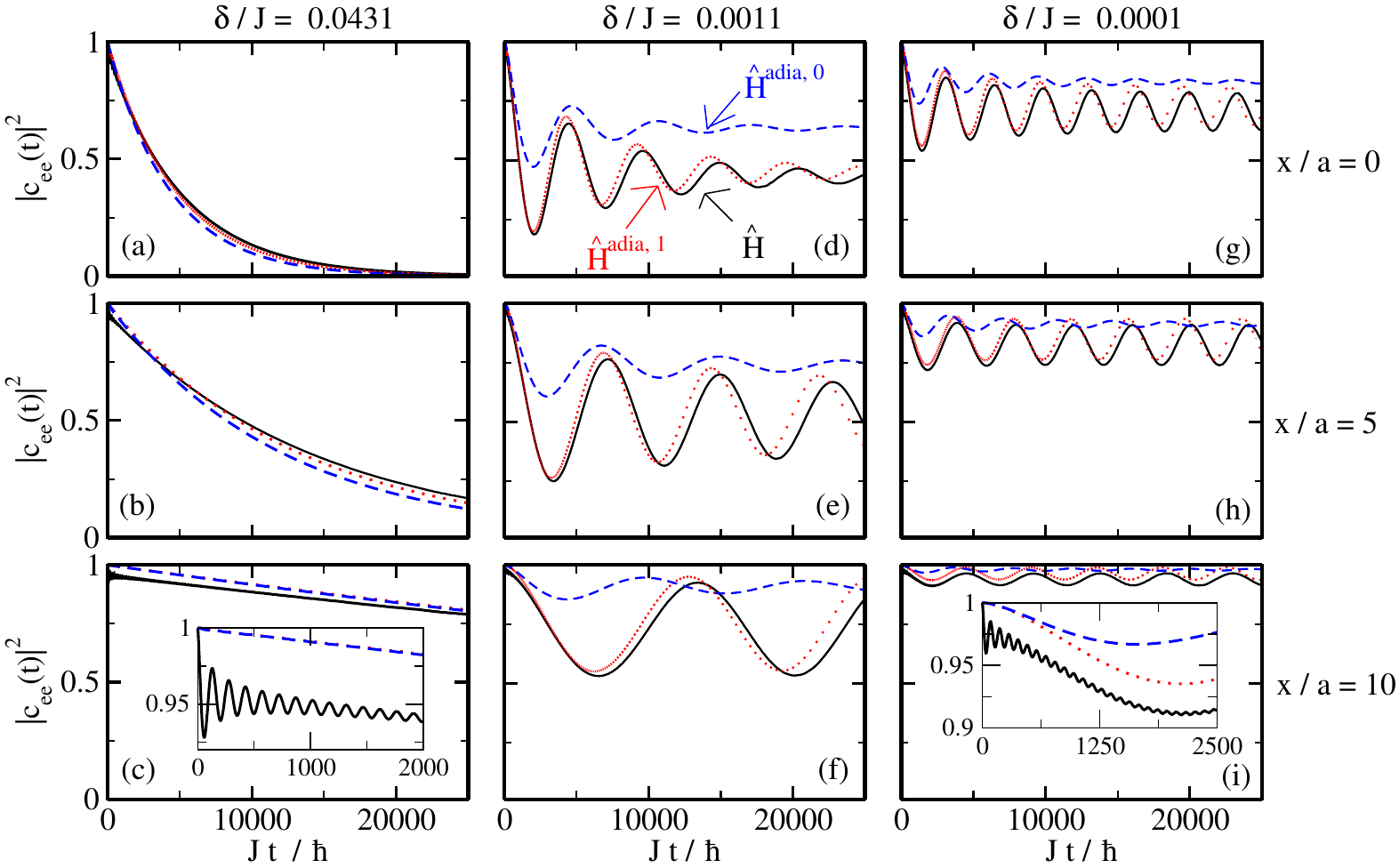}
\caption{Radiation dynamics for the initial state 
$|e,e,\text{vac} \rangle$, $U/J=-1$, and $g/J=1/50$. The lines show the population $P_{ee}(t)$ for various $x/a$ and $\delta/J$. 
The value of $x/a$ increases from the top row to the bottom row ($x/a=0$, $5$, and $10$ for the first,
second, and third row, respectively).
The value of $\delta/J$ decreases from the left most to the right most column ($\delta/J=0.0431$, $0.0011$, and $0.0001$ for the first,
second, and third column, respectively).
In all panels, the solid, dotted, and dashed lines show $P_{ee}(t)$ obtained by propagating the initial
state $|e,e,\text{vac} \rangle$ under the Hamiltonian $\hat{H}$, $\hat{H}^{\text{adia},1}$, and 
$\hat{H}^{\text{adia},0}$, respectively.
The data shown in panels (a)-(f) are also shown in Ref.~\protect\cite{ref_jugal}.
}
\label{fig_radiation_dynamics}
\end{figure}    

\end{widetext}

Figure~\ref{fig_radiation_dynamics} shows the time evolution of 
the population $P_{ee}(t)$
of state $|e,e,\text{vac} \rangle$
for $U/J=-1$.
For $\delta/J=0.0431$ (left column),
$P_{ee}(t)$ falls off 
roughly exponentially.
The decay is faster for $x/a=0$ [Fig.~\ref{fig_radiation_dynamics}(a)]
than for 
$x/a=5$
and $x/a=10$ 
[Figs.~\ref{fig_radiation_dynamics}(b)
and \ref{fig_radiation_dynamics}(c)].
For this large $\delta/J$, the
Markov approximation works well
and the agreement between
the results for
$\hat{H}$ (solid lines), $\hat{H}^{\text{adia,1}}$ (dotted lines),
and $\hat{H}^{\text{adia,0}}$ (dashed lines) is
quite good.
For $x/a=0$, $5$, and $10$, the population 
of the single-photon states (i.e., the sum of $\sum_{j=1,2}\sum_k|c_{jk}|^2 $)
is approximately equal to $1.8$~\%, $2.0$~\%, and $4.0$~\%, respectively. 
The comparatively large population of the single-photon states for $x/a=10$ signals that the adiabatic elimination deteriorates for large $x/a$.
The inset of Fig.~\ref{fig_radiation_dynamics}(c) shows that the radiation emitted by the first 
and second emitters are uncorrelated 
initially. We find that the oscillations
displayed by the black solid
line are well reproduced by the single-emitter dynamics, i.e., by treating the two emitters as independent quantities (effectively, this corresponds to setting $U=0$).

For small $\delta/J$, the dynamics changes significantly. The middle  and right most columns of Fig.~\ref{fig_radiation_dynamics} correspond to $\delta/J=0.0011$ and $\delta/J=0.0001$, respectively.
For these two detunings, the population $P_{ee}(t)$ does not change exponentially but instead
exhibits damped or essentially undamped oscillatory
behaviors for $x/a=0$, $5$, and $10$.
For all six parameter combinations [Figs.~\ref{fig_radiation_dynamics}(d)-\ref{fig_radiation_dynamics}(i)],
the adiabatic elimination Hamiltonian $\hat{H}^{\text{adia,1}}$ (red dotted lines), which accounts for the effective interactions $\underline{G}_{K,K'}(n_1,n_2)$, reproduces the key features of the dynamics of the full Hamiltonian $\hat{H}$ (black solid line)---such as the amplitude, frequency, and degree of damping---faithfully. 
The adiabatic elimination Hamiltonian $\hat{H}^{\text{adia,0}}$ (blue dashed lines), in contrast, provides a comparatively poor description of the oscillatory dynamics
[Figs.~\ref{fig_radiation_dynamics}(d)-\ref{fig_radiation_dynamics}(i)].
The comparison shows that appearance of essentially undamped oscillations depends critically on the effective interactions $\underline{G}_{K,K'}(n_1,n_2)$; recall, these are not
included in $\hat{H}^{\text{adia,0}}$. 
The inset of Fig.~\ref{fig_radiation_dynamics}(i) illustrates, as for the larger detuning, that the elimination of the single-photon states from the Hilbert space does remove fast oscillations and fails to capture the initial decay of $P_{ee}(t)$ that is due to uncorrelated decay of single photons.

\begin{figure}
\includegraphics[width=0.47\textwidth]{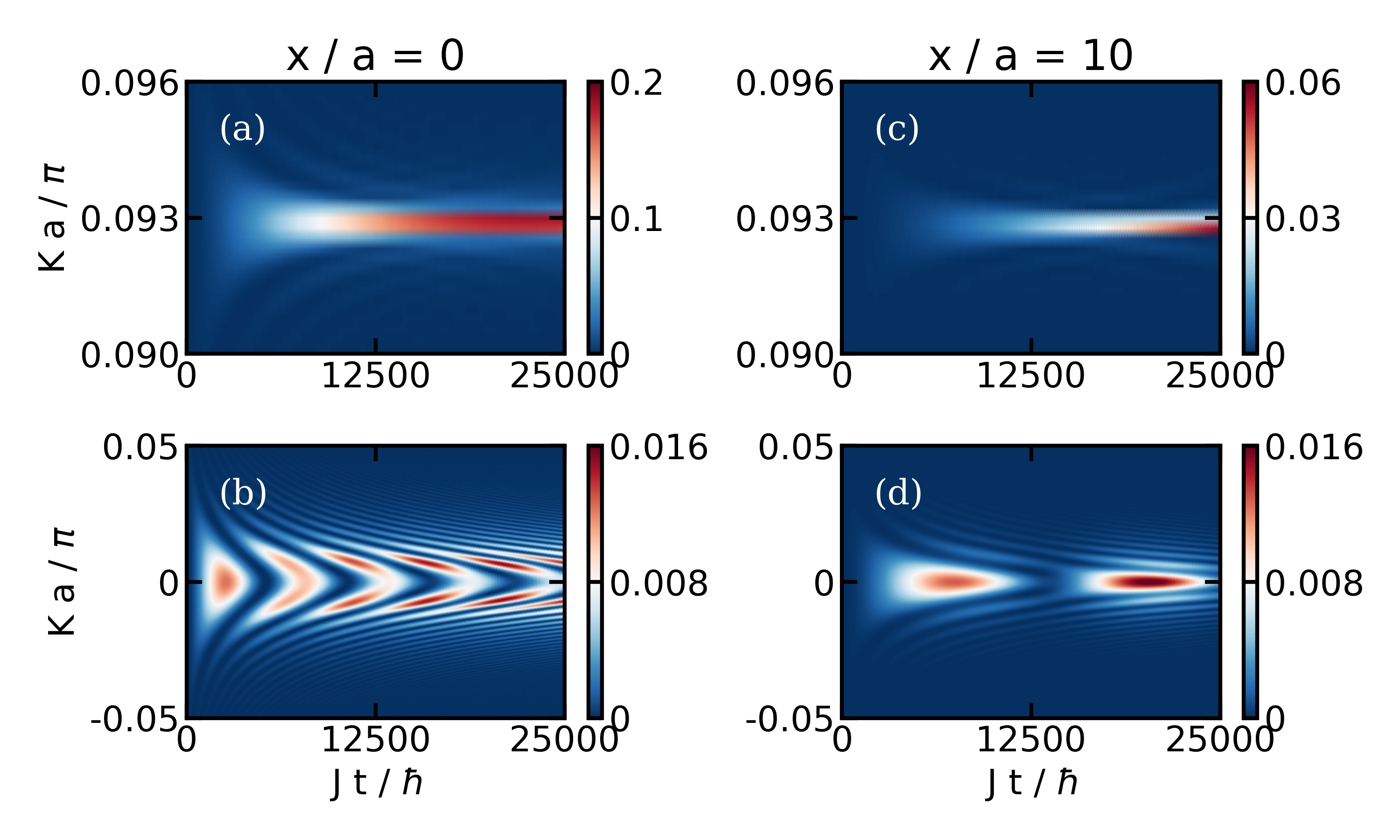}
\caption{Contour plot of the populations $|c_{K,b}(t)|^2$ of the two-photon bound states $|g,g,K\rangle$ as functions of the dimensionless center-of-mass wave number $Ka/ \pi$ and the dimensionless time $Jt / \hbar$ for $U/J=-1$ and $g/J=1/50$. (a) $\delta/J=0.0431$ and $x/a=0$. 
(b) $\delta/J=0.0011$ and $x/a=0$. (c) $\delta/J=0.0431$ and $x/a=10$. (d) $\delta/J=0.0011$ and $x/a=10$. The color scheme and range of the vertical axis are adjusted in each panel for ease of viewing.}
\label{fig_cK-dynamics}
\end{figure}

Figure~\ref{fig_cK-dynamics} shows the populations $|c_{K,b}(t)|^2$ of the two-photon states
$|g,g,K \rangle $
as functions of $J t/\hbar$ and $Ka/\pi$
for $\delta/J=0.0431$ (top row) and
$\delta/J=0.0011$ (bottom row).
The behavior for large and small detunings is distinct. For $\delta/J=0.0431$, a few $Ka/\pi$ values---centered around $K^{(*)}a/\pi$---get populated as time increases for $x/a=0$ [Fig.~\ref{fig_cK-dynamics}(a)] and $x/a=10$ [Fig.~\ref{fig_cK-dynamics}(c)]. The excitations, which exist initially in the form of matter, get transferred to the photons. Since the decay involves multiple states, the radiation emitted is incoherent. For $\delta/J=0.0011$, the populations $|c_{K,b}(t)|^2$ with $K\approx 0$ oscillate in time for $x/a=0$ [Fig.~\ref{fig_cK-dynamics}(b)] and $x/a=10$ [Fig.~\ref{fig_cK-dynamics}(d)]. As expected, the oscillation frequencies are the same as those displayed in Figs.~\ref{fig_radiation_dynamics}(d) and \ref{fig_radiation_dynamics}(f).

The undamped Rabi oscillations displayed in Fig.~\ref{fig_radiation_dynamics}
are readily explained by the fact that 
the Hamiltonian $\hat{H}$ supports two bound states for
sufficiently large $x/a$. The initial state can, to a good approximation be written as a superposition of
the symmetric and anti-symmetric hybridized energy eigen states $\Psi_+$ and $\Psi_-$. As a function of time,
population is transferred between the two bound energy eigen states, with the angular oscillation frequency being
equal to 
$(E_--E_+) / \hbar$.

To explain the damping,
we decompose the initial state $|e,e,\text{vac} \rangle$ into the energy eigen states $\phi_E$ of $\hat{H}$. For this calculation, we divide the energy eigen states into
two groups. The state $\phi_{0}$ with energy $E_0$ (lowest energy eigen state) and the states $\{ \phi_{j}\}$ with energy $E_j$ ($j=1,2,\cdots$; all other states). The latter group of states includes the scattering states and the hybridized state $\Psi_-$, whose energy is either just below or immersed into the scattering continuum. Using this grouping, we find
\begin{eqnarray}
\label{eq_damping1}
|c_{ee}(t)|^2 
\approx
(P_{0})^2 + 2 P_{0} \sum_{j>0}  P_j \cos \left[\frac{(E_0-E_j)t}{\hbar} \right] ,
\end{eqnarray}
where the time-independent probabilities $P_j$ are given by
\begin{eqnarray}
P_j = | \langle e,e,\text{vac}| \phi_j \rangle|^2
\end{eqnarray}
(the $P_j$ are positive).
In writing Eq.~(\ref{eq_damping1}),
we dropped the term $C(t)$,
\begin{eqnarray}
C(t)= \sum_{j>0,j'>0} P_j P_{j'} \cos \left[\frac{(E_j-E_{j'})t}{\hbar} \right],
\end{eqnarray}
on the right hand side;
we find numerically that the term $C(t)$
contributes neglegibly to $|c_{ee}(t)|^2$.
The damping of the Rabi oscillations is thus due to  the energy spread of the energy states $\phi_j$ ($j>0$) with non-vanishing $P_j$. We refer to this as  dephasing.

\begin{figure}
\includegraphics[width=0.30\textwidth]{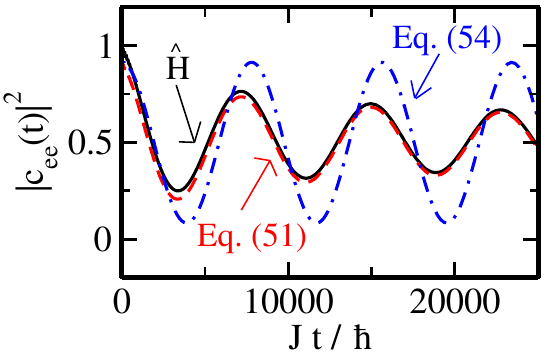}
\caption{Population $P_{ee}(t)=|c_{ee}(t)|^2$ as a function of time for $U/J=-1$, $g/J=1/50$, $\delta/J=0.0011$, and $x/a=5$. The black solid line shows $P_{ee}(t)$ obtained by propagating the initial
state $|e,e,\text{vac} \rangle$ under the Hamiltonian $\hat{H}$. The red dashed and blue dash-dotted lines show $P_{ee}(t)$ obtained using  Eqs.~(\ref{eq_damping1}) and (\ref{eq_simple_average}), respectively.
}
\label{fig_SA-dephasing}
\end{figure}

If we replace the energies $E_j$ with $j>0$ in Eq.~(\ref{eq_damping1}) by
$E_-$,
we find
\begin{eqnarray}
\label{eq_simple_average}
|c_{ee}(t)|^2 \approx (P_{0})^2 + 2 P_{0} (1-P_0) \cos \left[\frac{(E_0-E_-)t}{\hbar} \right].
\end{eqnarray}
The fractional population~\cite{ref_fractional-pop}, i.e., the large $t$ limit of $|c_{ee}(t)|^2$, is 
to a very good approximation given by $(P_0)^2$.
Equation~(\ref{eq_simple_average}) describes undamped Rabi oscillations, which reproduce the short-time amplitude and oscillation frequency very well (see Fig.~\ref{fig_SA-dephasing}).
The damping, which is due---as already pointed out above---to the energy spread of the $E_j$ with $j>1$, is not captured
by Eq.~(\ref{eq_simple_average}).
Alternatively, the damping can be explained by using that the hybridized state $\Psi_-$, supported by $\hat{H}$ and $\hat{H}^{\text{adia,1}}$, is for small $x/a$ immersed in the scattering continuum. As such, its energy acquires an imaginary part, which provides a finite lifetime or damping coefficient.

\section{Conclusion}
\label{sec_conclusion}
This paper discussed the dynamics of two emitters coupled to a wave guide with Kerr-like non-linearity.
Our interest was in the regime where the two emitters are in resonance with the two-photon bound state supported by the 
one-dimensional wave guide. Even though the emitters are not interacting with each other, correlated dynamics is introduced through the coupling of the emitters to the wave guide.
The induced correlations occur on length scales that are comparable to the size of the two-photon bound state.
Somewhat surprisingly, a regime where the excitations are transferred back and forth between the emitter and photonic degrees of freedom is observed. The essentially undamped Rabi oscillations are due to the emergence two hybridized bound states 
whose energies lie in the band gap.
This behavior is unique to the two-emitter system: the single-emitter system does not display an analogous behavior.

Throughout, we worked in the weak coupling regime; specifically, the figures all use a coupling strength of $g/J=1/50$.
In the Markovian regime [scenario (A);
see Fig.~\ref{fig1}(b)], the coupling constant enters as a multiplicative factor, i.e., the decay constant $\Gamma_{\text{bath}}$ is
directly proportional to $(g/J)^4$ [see Eq.~(\ref{eq_gammabath})].
This regime had previously been investigated in Ref.~\cite{ref_rabl_non-linear}.
For resonance wave vectors $K^{(*)}$ near the bottom of the band [scenarios (B) and (C); see Fig.~\ref{fig1}(b)], the 
$g$-dependence is more intricate. Within the two-state Hamiltonian $\hat{H}^{\text{2-st.}}$, the coupling constant enters through $\Delta_e$, $E_{\text{pol}}$, and $G_{\text{eff}}(n_1,n_2)$:
$\Delta_e$ is directly proportional to $-(g/J)^2$,
$E_{\text{pol}}$ contains a term that is proportional to $-(g/J)^4$,
and
$G_{\text{eff}}(n_1,n_2)$ is directly proportional to $(g/J)^3$.
Because of this non-trivial
$g$-dependence, the energies of the hybridized eigen states and, correspondingly, the Rabi oscillation frequency vary notably with $g/J$.
In addition, the regime where the two-state Hamiltonian $\hat{H}^{\text{2-st.}}$ provides a reliable description depends on $g/J$. For $g/J$ values that are smaller than the value considered in this paper, the observation of Rabi oscillations requires smaller detuning $\delta/J$.
Conversely, a larger $g/J$ allows for the observation of Rabi oscillations for larger $\delta/J$. It is an open question how large $g/J$ can be before counter-rotating terms, which are not included in $\hat{H}$, play a non-negligible role.
We are not aware of any previous work on cavity arrays with Kerr-like non-linearity coupled to two-level emitters that looked at parameter combinations corresponding to scenarios~(B) and (C).
As shown in this paper, these scenarios give rise to qualitatively new behaviors that are inaccessible in the absence of the non-linearity and in cavity array--single-emitter systems.

Our treatment neglects, as already pointed out in the last paragraph of Sec.~\ref{sec_hamiltonian},
single-photon losses. If the single-photon loss rate is denoted by $\kappa$, the exponential decay for the initial state $|e,\text{vac} \rangle$
is characterized by $\Gamma_1$, where $\Gamma_1=\kappa g^2/[4 J^{1/2} (\hbar \omega_c -2J- \hbar \omega_e )^{3/2}]$~\cite{ref_rabl_atom-field}. For the dynamics to be dominated by correlated two-photon processes, we
must thus require $\Gamma_1 \ll \Gamma_{\text{bath}}$ or, dropping all factors that are (roughly) of order $1$, $\hbar \kappa/J \ll (g/J)^2$.
Reference~\cite{ref_rabl_non-linear} argues that this regime can be reached with non-linear photonic lattices or superconducting qubits coupled to an array of microwave resonators.
Recent experimental work on two transmon qubits coupled to a superconducting microwave photonic crystal, e.g.,  demonstrated tunable onsite and interbound state interactions~\cite{ref_houck}.

The results presented open the door for several follow-up investigations.
Continuing to work in the two-excitation sub-space, it would be interesting to consider an emitter array coupled to the non-linear wave guide. Intriguing hopping dynamics of the radiation might be observed when the radiation is initially localized in two of the emitters. In addition, it might be interesting to investigate the dependence of the dynamics on the initial state. For example, it might be interesting to compare the dynamics for initial states that can be written as a product states to that for entangled superposition states.

{\em{Acknowledgement:}}
Support by the National Science Foundation through
grant number
PHY-1806259 is
gratefully acknowledged.
This work used
the OU
Supercomputing Center for Education and Research
(OSCER) at the University of Oklahoma (OU).

\appendix

\section{Single-emitter dynamics}
\label{appendix_single_emitter}

Section~\ref{sec_results} uses
the dynamics of a single emitter coupled to cavity $n$ as a reference.
This appendix summarizes the single-emitter results. Note that
the results are independent of $n$ since the emitter position does not matter.

 We expand the time-dependent wave packet as~\cite{ref_marknonmark,ref_fractional-pop}
\begin{eqnarray}
|\psi (t) \rangle &=& \exp(-\imath \omega_e t) \times \nonumber \\
&&\left[
d_e(t) |e,\mbox{vac} \rangle + \sum_k d_k(t) \hat{a}_k^{\dagger} |g, \mbox{vac} \rangle
\right],
\end{eqnarray}
where $d_e(t)$ and $d_k(t)$ are expansion coefficients. 
Starting at time $t=0$ in the state $|e,\text{vac}\rangle$ [i.e., setting $d_e(0)=1$ and $d_k(0)=0$] and following the steps of Ref.~\cite{ref_fractional-pop}, one finds
\begin{eqnarray}
\label{eq_single_emitter_coeff}
\dot{d}_e(t) = 
- \int_0^t
d_e(t-t') {\cal{M}}(t')  dt', 
\end{eqnarray}
where 
\begin{eqnarray}
{\cal{M}}(t')=
\frac{g^2}{\hbar^2 N} \sum_k  
\exp \left(
-\frac{\imath {\Delta}_{k} t'}{\hbar}
\right).
\end{eqnarray}
The integral in Eq.~(\ref{eq_single_emitter_coeff}) can be evaluated analytically~\cite{ref_marknonmark}.

In what follows, we review results obtained within the Markov approximation~\cite{ref_fractional-pop,ref_marknonmark}. The presence of the bath memory function ${\cal{M}}(t')$
in Eq.~(\ref{eq_single_emitter_coeff}) 
indicates that the dynamics is, in general, non-Markovian: the evolution of the coefficient
${d}_{e}(t)$ depends on the past, i.e., the system's state at earlier times.
If the bath memory time $\tau_{\text{bath}}$ is short, i.e., if
\begin{eqnarray}
\left| \frac{\dot{{d}}_{e}(t)}{{d}_{e}(t)} \tau_{\text{bath}} \right|
 \ll 1,
\end{eqnarray}
the Markov approximation 
\begin{eqnarray}
\int_0^{t}  {\cal{M}}(t') {d}_{e}(t-t') dt' \approx 
{d}_{e}(t) \int_0^t  {\cal{M}}(t') dt'
\end{eqnarray}
should be reliable.
Using 
\begin{eqnarray}
\text{Re} \left[ \lim_{t \rightarrow \infty} \int_0^t \exp \left( - \frac{\imath {\Delta}_{k} t'}{\hbar} \right) dt' \right] 
= \pi \hbar \delta( {\Delta}_{k})
\end{eqnarray}
and dropping the imaginary part, which introduces a negligible energy shift $\Delta_{\text{single}}$,
Eq.~(\ref{eq_single_emitter_coeff}) can be rewritten as
\begin{eqnarray} 
\label{eq_appendix_diff_single}
\dot{{d}}_{e}(t) = -\Gamma_{\text{single}}  {d}_{e}(t)
\end{eqnarray}
or
\begin{eqnarray}
d_e(t) =\exp 
 \left( -\Gamma_{\text{single}} t \right),
 \end{eqnarray}
where
\begin{eqnarray}
\label{eq_gammabath0_s}
\Gamma_{\text{single}}= \frac{\pi g^2}{\hbar  N} \sum_k \delta({\Delta}_{k}).
\end{eqnarray}
Equation~(\ref{eq_appendix_diff_single}) describes the effect of the bath on the expansion coefficient of the state $| e, \text{vac}\rangle$. The emitter-bath coupling induces an exponential decay of the population, with a decay rate $2\Gamma_{\text{single}}$, and an energy shift $\Delta_{\text{single}}$ in the energy of the state $| e, \text{vac}\rangle$.

To find an explicit expression for $\Gamma_{\text{single}}$, we replace the sum over $k$ by an integral,
\begin{eqnarray}
\label{eq_sumintegral_s}
 \sum_k  (\cdots)
= \frac{Na}{2 \pi} \int_{-\pi/a}^{\pi/a} 
(\cdots) dk,
\end{eqnarray}
and perform a change of variable,
\begin{eqnarray}
d k =\frac{\partial k}{\partial {E}_{k}} d  {E}_{k}=
\left( \frac{\partial E_{k}}{\partial k } \right)^{-1}d  {E}_{k}.
\end{eqnarray}
Defining the density of states $\rho_{\text{single}}(k)$ through
\begin{eqnarray}
\label{eq_dos_s}
\rho_{\text{single}}(k)=
J \left( \frac{\partial E_{k}}{\partial k} \Large  \right)^{-1}
\end{eqnarray}
and using Eqs.~(\ref{eq_sumintegral_s})-(\ref{eq_dos_s}) in Eq.~(\ref{eq_gammabath0_s}),
we find 
\begin{eqnarray}
\Gamma_{\text{single}} = 
\frac{g^2 a}{ \hbar J} \int_{\hbar \omega_c-2J}^{\hbar \omega_c+2J} \delta(E_k-\hbar \omega_e) \rho_{\text{single}}(k) d E_k.
\end{eqnarray}
Evaluating the integral yields Eq.~(\ref{eq_gammasingle}) from the main text.

\section{Details on adiabatic elimination}
\label{appendixA}
Equations~(\ref{eq_coeffe})-(\ref{eq_coeffscatt}) of the main text 
are equivalent to 
the time-dependent Schr\"odinger equation within the two-excitation subspace. After adiabatic elimination, the equations reduce to
\begin{widetext}
\begin{eqnarray}
\label{eq_appendixA_first}
\imath \hbar \dot{c}_{ee}(t) = 
2 \Delta_e c_{ee}(t)
&+& \frac{g^2}{J\sqrt{N}}
\sum_K
F_{K,b}(n_1,n_2)  c_{K,b}(t) + \frac{g^2}{J\sqrt{N}} \sum_K
\sum_q F_{K,q}(n_1,n_2)  c_{K,q}(t) 
,
\end{eqnarray} 
\begin{eqnarray}
\label{eq_appendixA_first2}
\imath \hbar \dot{c}_{K,b}(t) = 
\Delta_{K,b} c_{K,b}(t) +
\frac{g^2}{J\sqrt{N}}  
F_{K,b}^*(n_1,n_2) c_{ee}(t)+
\frac{g^2}{JN} \sum_{K', K} G_{K, K'}(n_1,n_2) c_{K',b}(t)+ \nonumber \\
\frac{g^2}{JN} \sum_{K'}\sum_q
H_{K,K',q}(n_1,n_2) c_{K',q}(t), 
\end{eqnarray}
and
\begin{eqnarray}
\label{eq_appendixA_first3}
\imath \hbar \dot{c}_{K,q}(t) = 
\Delta_{K,q} c_{K,q}(t) +\frac{g^2}{J\sqrt{N}} F_{K,q}^*(n_1,n_2) c_{ee}(t) +\frac{g^2}{JN} \sum_{K' , K} \sum_{q' , q} [H_{K',K,q'}(n_1, n_2)]^*  c_{K',b}(t) + \nonumber \\
\frac{g^2}{JN} \sum_{K' , K} \sum_{q' , q} J_{K,K',q,q'}(n_1, n_2)  c_{K',q'}(t) ,
\end{eqnarray}
where 
\begin{eqnarray}
\label{eq_fsubcapk}
F_{K,b}(n_1,n_2)=-
\sum_k \sum_{\alpha=1,2}
\frac{J}{N\Delta_k}
\exp( -\imath k a n_{\beta(\alpha)} ) 
[M_b(k, n_{\alpha},K)]^*,
\end{eqnarray}
\begin{eqnarray}
\label{eq_fsubcapq}
F_{K,q}(n_1,n_2)=-
\sum_k  \sum_{\alpha=1,2}
\frac{J}{N\Delta_k}
\exp( -\imath k a n_{\beta(\alpha)} )
 [M_q(k, n_{\alpha},K)]^*,
\end{eqnarray}
\begin{eqnarray}
\label{eq_capg}
G_{K,K'}(n_1,n_2) =
-
 \sum_k \sum_{\alpha=1,2}
\frac{J}{N\Delta_k}
[M_b(k,n_{\alpha},K)]^* M_b(k, n_{\alpha},K'),
\end{eqnarray}
\begin{eqnarray}
H_{K,K',q}(n_1,n_2) = -
\sum_k \sum_{\alpha=1,2}
\frac{J}{N\Delta_k}
[M_b(k,n_{\alpha},K)]^* M_q(k, n_{\alpha},K'),
\end{eqnarray}
\begin{eqnarray}
J_{K,K',q,q'}(n_1,n_2) = 
-\sum_k \sum_{\alpha=1,2} \frac{J}{N\Delta_k}
[M_q(k,n_{\alpha},K)]^* M_{q'}(k,n_{\alpha},K'),
\end{eqnarray}
\end{widetext}
and
$\Delta_e$ is given in Eq.~(\ref{eq_appendixA_last}) of the main text.
The quantity $2\Delta_e$ can be interpreted as an effective Stark shift~\cite{ref_rabl_non-linear} 
that is introduced by the single-photon states. Before the adiabatic elimination, energies are measured relative to 
the energy $2 \hbar \omega_e$ of the initial state. After the adiabatic elimination, the state with two emitter
excitations is ``detuned with respect to itself".
The Stark shift was set to zero in Ref.~\cite{ref_rabl_non-linear}. This approximation is justified when
the resonance wave vector lies in the middle of the band. Near the bottom of the band, however, the quantity
$2 \Delta_e$ introduces
a non-perturbative correction~\cite{ref_jugal}. 

The quantity $G_{K,K'}(n_1, n_2)$ describes  effective off-diagonal couplings between 
two-photon bound states with center-of-mass wave numbers $K$ and $K'$. Before the adiabatic
elimination, the right hand side of the equation for
$\dot{c}_{K,b}(t)$ does not depend on $c_{K',b}(t)$ for $K' \ne K$.
After the adiabatic elimination, the right hand side of the equation for
$\dot{c}_{K,b}(t)$ depends on $c_{K',b}(t)$ for $K' \ne K$.

The equations above simplify significantly if 
the
contributions from the scattering states 
are dropped. The result is given in Eqs.~(\ref{eq_se_adia1})-(\ref{eq_se_adia4})
of the main text.
The next step is to go to a rotating frame.
Rewriting the coupled equations in terms of $\tilde{c}_{ee}(t)$ and $\tilde{c}_{K,b}(t)$,
where 
\begin{eqnarray}
\label{eq_coeffe_tilde}
\tilde{c}_{ee}(t)= \exp \left( 2 \imath \Delta_e t / \hbar \right) c_{ee}(t)
\end{eqnarray}
and
\begin{eqnarray}
\tilde{c}_{K,b}(t)= \exp \left(  \imath \Delta_{K,b} t / \hbar \right) c_{K,b}(t),
\end{eqnarray}
the
diagonal terms vanish:
\begin{eqnarray}
\label{eq_coeffe_tilde4}
\imath \hbar \dot{\tilde{{c}}}_{ee}(t) = \nonumber \\
\frac{g^2}{J\sqrt{N}}
\sum_K
F_{K,b}(n_1,n_2)
\exp \left( - \frac{\imath  {\tilde{\Delta}_{K,b}} t }{ \hbar} \right)
\tilde{c}_{K,b}(t) 
\end{eqnarray} 
and 
\begin{eqnarray}
\label{eq_coeffbound_tilde4}
\imath \hbar \dot{\tilde{c}}_{K,b}(t) = \nonumber \\
\frac{g^2}{J\sqrt{N}}  [F_{K,b}(n_1,n_2)]^* 
\exp \left( \frac{\imath {\tilde{\Delta}_{K,b}}  t }{ \hbar} \right)
\tilde{c}_{ee}(t)
\end{eqnarray}
with
\begin{eqnarray}
\tilde{\Delta}_{K,b} = \Delta_{K,b} - 2 \Delta_e.
\end{eqnarray}

We now specialize to
the initial state
$|e,e,\mbox{vac} \rangle$. Integrating Eq.~(\ref{eq_coeffbound_tilde4})
and using the result in Eq.~(\ref{eq_coeffe_tilde4}),
we obtain an equation for the coefficient
$\tilde{c}_{ee}(t)$ that is independent of $\tilde{c}_{K,b}(t)$: 
\begin{eqnarray}
\label{eq_coeffe_tilde2}
\dot{\tilde{c}}_{ee}(t) = 
-\int_0^{t}  {\cal{M}}(t',n_1,n_2) \tilde{c}_{ee}(t-t') dt', 
\end{eqnarray}
where the bath memory function ${\cal{M}}(t',n_1,n_2)$ reads
\begin{eqnarray}
{\cal{M}}(t',n_1,n_2)= \nonumber \\
\frac{g^4}{\hbar^2 J^2 N} \sum_K |F_{K,b}(n_1,n_2)|^2 
\exp \left(
-\frac{\imath \tilde{\Delta}_{K,b} t'}{\hbar}
\right).
\end{eqnarray}

Following the same steps as in Appendix~\ref{appendix_single_emitter}
and
defining the density of states $\rho(K)$ through
\begin{eqnarray}
\label{eq_dos}
\rho(K)=
J \left( \frac{\partial \Delta_{K,b}}{\partial K} \Large  \right)^{-1},
\end{eqnarray}
we find 
Eq.~(\ref{eq_gammabath}) of the main text.


\begin{thebibliography}{100}
\bibitem{footnote_detuning} 
Note that our
definition of the detuning $\delta$ differs from that employed in Ref.~\cite{ref_rabl_non-linear}.
\end{thebibliography}
\end{document}